\documentclass[11pt]{article}

\usepackage[utf8]{inputenc}
\usepackage[english]{babel}
\usepackage[normalem]{ulem}
\usepackage[dvips]{epsfig}
\usepackage[title]{appendix}
\usepackage{bbm}
\usepackage{amsmath,amssymb,latexsym,amscd}
\usepackage{amsthm}
\usepackage{graphicx}
\usepackage{multirow, array}
\usepackage{verbatim}
\usepackage{float}
\usepackage{xargs}    
\usepackage{upgreek}

\usepackage{dsfont}

\usepackage{mathtools}

\usepackage{anysize}
\marginsize{2cm}{2cm}{2cm}{2cm}

\usepackage{enumerate}%%%% Enumerar \begin{enumerate}[(a)] % (a), (b), (c), ...
\usepackage{mathrsfs}
\usepackage{amsfonts}
\usepackage{hyphenat}
\usepackage{bbm}
\usepackage{amssymb}
\usepackage{amsthm}
\usepackage{xparse}
\usepackage{physics}
\usepackage{hyperref}
\usepackage{accents}
\usepackage{textgreek}
\usepackage[scr=boondox]{mathalfa}
\usepackage{dirtytalk}
\usepackage{verbatim}
\usepackage{color}
\newcommand{\R}{\mathbb R}
\newcommand{\N}{\mathbb N}
\newcommand{\Z}{\mathbb Z}

\newcommand{\C}{\mathbb C}
\newcommand{\Ind}{\mathbbm{1}}
\newcommand{\fl}{\bigtriangleup}

\usepackage[left=4 cm,top=4 cm,right=2.5cm,bottom=2.5cm]{geometry}

\usepackage{comment}
\newcounter{thm}
\newtheorem{theorem}[thm]{Theorem}
\newtheorem{proposition}[thm]{Proposition}

\newtheorem{lemma}[thm]{Lemma}

\newtheorem{definition}[thm]{Definition}

\numberwithin{equation}{section}

\title{Decay of the Green's function of the fractional Anderson model and connection to long-range SAW}
\author{Margherita Disertori\footnote{University of Bonn, Institute for Applied Mathematics  \&
Hausdorff Center for Mathematics,Endenicher Allee 60, 53115 Bonn, Germany. E-mail:disertori@iam.uni-bonn.de},  Roberto Maturana Escobar\footnote{University of Bonn, Institute for Applied Mathematics  \&
Hausdorff Center for Mathematics,Endenicher Allee 60, 53115 Bonn, Germany. E-mail:maturana@iam.uni-bonn.de}, 
 Constanza Rojas-Molina\footnote{Laboratoire AGM, D\'epartement de Math\'ematiques,CY Cergy Paris Universit\'e, 2 Av. Adolphe Chauvin, 95302 Cergy-Pontoise, France. E-mail:crojasmo@cyu.fr},}
\date{\today}
\begin{document}
\maketitle
\begin{abstract} We prove a connection between the Green's function of the fractional Anderson model and the two point function
of a self-avoiding random walk with long range jumps, adapting a strategy proposed by Schenker in 2015.
This connection allows us to  exploit results from the theory of  self-avoiding random walks to improve previous bounds known
for the  fractional Anderson model at strong disorder. In particular, we enlarge the range of the disorder parameter
where spectral localization occurs. Moreover we prove that the decay of Green's function  at strong disorder
for any $0<\upalpha <1$ is arbitrarily close to the decay of  the massive resolvent of the corresponding
fractional Laplacian, in agreement with the case of the standard Anderson model $\upalpha =1$.
We also derive upper and lower bounds for the resolvent of the discrete fractional Laplacian 
with arbitrary mass $m\geq 0,$ that are of independent interest.\vspace{0,2cm}

\vspace{0,2cm}
 	

\end{abstract}

\section{Introduction }\label{Introduction}
Transport phenomena in disordered environment are often described via
random Schr\"o- dinger
operators. On the lattice $\Z^{d},$ $d\geq 1,$ they take the
form of an infinite random matrix
$H_{\omega }= T+\lambda V_{\omega }\in
 \R^{\Z^{d}\times \Z^{d}}_{sym}$ where $T$ is a deterministic matrix (the kinetic part) and $V_{\omega }$ is a diagonal matrix with random entries.
In its most standard formulation $T$  is the negative discrete Laplacian  $-\Delta $
defined via $-\Delta (x,y):= -\delta_{|x-y|=1}+2d\,\delta_{|x-y|=0}, $
where $|\cdot|$ denotes the $\ell^{2}$ norm. This defines a self-adjoint bounded operator
$-\Delta\colon \ell^2(\Z^{d})\to \ell^2(\Z^{d})$ with absolutely continuum spectrum and delocalized generalized eigenfunctions.
More generally, $T$ can be a symmetric matrix with decaying off-diagonal terms. 

Operators of the form $H=T+V$ where $V$ is a, possibly random, multiplication operator and $T$ is long-range  
have attracted increasing interest in recent years
\cite{Han2019,Padgett_Liaw_etal19,gebert2020lipschitz,shamis2021upper,Jitomirskaya-Liu,Liu_2023,shi2023localization,shi21}.
In particular, the usual exponential decay of eigenfunctions and dynamical bounds is replaced in this case by a polynomial decay.

In this paper we consider the case when $T$ is  the discrete fractional Laplacian
$(-\Delta)^\upalpha$  with $0<\upalpha <1,$ which is obtained from $-\Delta$ via functional calculus. This operator has been subject to increasing interest in recent years.
Just as the standard discrete Laplacian, it is bounded and  translation invariant
$(-\Delta)^\upalpha (x,y)=(-\Delta)^\upalpha (0,y-x)$, non-negative as a quadratic form 
and satisfies (see \cite[Thm. 2.2]{gebert2020lipschitz})
\[
(-\Delta)^\upalpha (x,x)>0, \qquad (-\Delta)^\upalpha (x,y)\leq 0 \quad \forall x\neq y.
\]
Its off-diagonal matrix elements decay polynomially. See  \cite[Thm.~2.2(iii)]{gebert2020lipschitz},
\cite[Lemma~2.1]{Slade2018}
or, for the one dimensional case, \cite[Thm.~1.1]{ciaurri2015fractional}. More precisely
there are constants $0<c_{\upalpha,d}<\mathcal{C}_{\upalpha,d}$ such that
\begin{align}\label{Eq.decay}
        \frac{c_{\upalpha,d}}{\abs{x-y}^{d+2\upalpha}} \leq -(-\Delta)^\upalpha(x,y)\leq \frac{\mathcal{C}_{\upalpha,d}}{\abs{x-y}^{d+2\upalpha}},\qquad \forall x,y\in\Z^{d},\: x\neq y.
    \end{align} 
In particular $(-\Delta)^\upalpha $ has a summable kernel
$\sum_{x\in \Z^{d}} |(-\Delta)^\upalpha (0,x) |<\infty$ $\forall 0<\upalpha\leq 1 $
and 
\begin{equation}\label{eq:deltasum}
(-\Delta)^\upalpha (0,0)= -\sum_{x\neq 0}  (-\Delta)^\upalpha (0,x).
\end{equation}
The long range nature of  $(-\Delta)^\upalpha $ for $\upalpha <1$ changes drastically
the behavior of the resolvent $( (-\Delta)^\upalpha+m^{2})^{-1}$ with $m> 0$.
While for $\upalpha =1$ the corresponding kernel decays exponentially,
we only have  polynomial decay for $\upalpha <1$. Precisely, for $\upalpha=1$ 
\begin{equation}
c\, e^{-m|x-y|} \leq (-\Delta+m^{2})^{-1} (x,y)\leq
C\, e^{-m|x-y|} \qquad \forall x\neq y,
\end{equation}
for some constants $c,C>0,$ depending on $d,m,$ while for $\upalpha <1$
\begin{equation}\label{eq:massiveresolv}
\frac{c_{1}}{|x-y|^{d+2\upalpha }}\  \leq\ ((-\Delta)^\upalpha +m^{2})^{-1} (x,y)\ \leq\ \frac{C_{1}}{|x-y|^{d+2\upalpha }}\qquad \forall   \  x\neq y,
\end{equation}
for some constants $c_{1},C_{1}>0,$ depending on $d,m,\upalpha. $ See \cite[Lemma 3.2]{Slade2018}  or Thm.~\ref{prop:decay-massive-resolv}  in Section~\ref{sect:fractionalLapl}  below.
The limit $m\downarrow 0$  is well defined for
$d>2\upalpha, $ $0<\upalpha\leq 1$ and behaves polynomially both for $\upalpha =1$ and for $\upalpha <1.$
See \cite[Sect.~2]{Slade2018} and references therein,  Thm.~\ref{decayGreenszero} in Section~\ref{sect:fractionalLapl} below, or, for the one-dimensional case,
\cite[Thm.~1.3]{ciaurri2015fractional}. Precisely 
\begin{equation}
 \frac{c_{2}}{|x-y|^{d-2\upalpha }}\leq ((-\Delta)^\upalpha)^{-1} (x,y)\leq
 \frac{C_{2}}{|x-y|^{d-2\upalpha }}\qquad \forall x\neq y.
\end{equation}
for some constants $c_{2},C_{2}>0,$ depending  on  $d,\upalpha .$
\vspace{0,2cm}

In this paper we consider the so-called  fractional Anderson model, which is obtained by  perturbing
the fractional Laplacian with a random diagonal matrix as follows
\begin{equation}
\label{eq:TheOperator}
 \mathrm{H}_{\upalpha,\omega}= (-\Delta)^\upalpha +\uplambda \mathrm{V}_\omega\in
 \R^{\Z^{d}\times \Z^{d}}
\end{equation}
where  $\lambda >0$ is the disorder parameter,  
 $\mathrm{V}_\omega (x,y):= \delta_{|x-y|=0}\ \omega_{x}  $ and  
 $\omega:=(\omega_x)_{x\in\Z^{d}} \in \R^{\Z^{d}}$ is a family of i.i.d. real random variables endowed with the Borel probability measure $\mathbb{P} := \bigotimes_{\Z^{d}}\mathrm{P}_{0},$
 with compactly supported  one site   probability measure $\mathrm{P}_{0}$.
 With these assumptions the operator $\mathrm{H}_{\upalpha,\omega}$ is bounded and self-adjoint.
 By translation invariance it is also ergodic and hence the spectrum $\sigma (\mathrm{H}_{\upalpha,\omega})$ is a.s. a deterministic bounded interval of $\R,$ that we denote by $\Sigma.$

The fractional Anderson model in the discrete setting is known to exhibit pure point spectrum with eigenfunctions decaying at least polynomially at strong disorder,  see \cite[Thm.~3.2]{aizenman1993localization}, and
to exhibit fractional Lifshitz tails \cite{gebert2020lipschitz}. Localization for this operator is not yet available in the continuous setting, therefore it is important to understand better the mechanism causing
pure point spectrum.

In this paper we contribute to these efforts by giving an alternative proof of spectral localization that exploits a connection to self-avoiding random walks (SAW), following  \cite{schenker2015large} (see also \cite{tautenhan} and \cite{hundertmark}).
This allows us to improve  known results,  such as the polynomial decay rate of the  Green's function and eigenfunctions
at strong disorder. In particular we enlarge the range of the disorder parameter where spectral localization occurs.
Moreover we prove that the decay of Green's function  at strong disorder for any $0<\upalpha <1$ is arbitrarily
close to the decay of  the massive resolvent \eqref{eq:massiveresolv} of the corresponding fractional Laplacian.
We conjecture that this is the optimal decay rate one can obtain by the Fractional Moment Method.
Note that the same kind of result holds in the case of the standard Anderson model $\upalpha=1$.

%%%%%%%%%%%%%%%%%%%%%%%%%%%%%%%%%%%%%%%%%%%%%%%%%%%%%%%%%%%%%%%%%%%%
\paragraph{Organization of the paper.} In Section~\ref{sectmainresults} we state our main results
and discuss connections with the existing literature. In Section~\ref{sectprelim} we introduce 
the regularity assumption we need on the random potential and the basic definitions to introduce self-avoiding random walks, including a known result on the decay of the two-point correlation function of a
SAW with long jumps. In
Section~\ref{sectionSAW} we establish a comparison between the decay of the averaged fractional resolvent of
our model
and the two-point correlation function of a particular SAW with long jumps, proving our main result. Finally, in
Section~\ref{sect:fractionalLapl} we complement our results by studying properties of the discrete fractional Laplacian.
This section might be of independent interest.

\section{Main results and discussion}
\label{sectmainresults}

In the following we assume $\mathrm{P}_{0}$ is absolutely continuous with respect to Lebesgue and 
 $\tau-$regular for some  $\tau\in (\frac{d}{d+2\upalpha },1)$ with $\tau-$constant
$ M_{\tau }(\mathrm{P}_0)$ (cf. Def.~\ref{Def.tau.reg.} below).
Note that the decay of $(-\Delta)^{\upalpha } $, Eq. \eqref{Eq.decay}, ensures that
\begin{equation}\label{eq:deltassummable}
\sum_{x\in \Z^{d}} |(-\Delta)^\upalpha (0,x) |^{s}<\infty\qquad \mbox{holds }
 \forall s\in \left(\frac{d}{d+2\upalpha },1\right]
\end{equation}
since  in that interval we have $s (d+2\upalpha )>d.$ In particular this holds for all $s$ in the non empty interval $(\frac{d}{d+2\upalpha },\tau )$
and hence, by the Fractional Moment Method \cite[Thm~3.2]{aizenman1993localization},
the spectrum of $\mathrm{H}_{\upalpha,\omega}$  for large  disorder $\lambda >\lambda_{AM} (s),$ with
\begin{equation}\label{lambdaAM}
 \lambda_{AM} (s):=  M_{\tau }(\mathrm{P}_0)^{\frac{1}{\tau }}\left(\frac{2\tau }{\tau -s}
\sum_{x\in \Z^{d}}| (-\Delta )^{\upalpha } (0,x) |^{s} \right)^{\frac{1}{s}}
\end{equation}
consists only of  pure point spectrum,
with random square summable eigenvectors.
In this sense the fractional Anderson model undergoes the same localization phenomenon at large disorder as the standard non fractional one, but contrary to the case $\upalpha =1$, the operator $\mathrm{H}_{\upalpha,\omega}$ is expected to undergo a phase transition in $d=1$ for $\upalpha <1/2$  and in $d=2$ for all $0<\upalpha<1 $
between complete pure point spectrum at large disorder and coexistence of absolutely continuous  and pure
point spectrum at weak disorder \cite{JM98}.
Indeed, in dimension $d=1$, the random walk with long jumps generated by $-(-\Delta)^\upalpha$ is transient in the case  $0<\upalpha<1/2,$ and recurrent in the case $1/2\leq \upalpha \leq1$, while
in dimension $d=2$ and above, it is transient for all $0<\upalpha<1$, see e.g. \cite[Appendix B.1]{CaputoFaggionatoGaudilliere09}.

The situation changes drastically when considering the spatial decay of the corresponding eigenvectors.
While these are exponentially localized around some random point for $\upalpha =1,$  polynomial decay
is expected   for $\upalpha <1$  in any dimension due to the long range nature of 
$(-\Delta)^\upalpha.$ Upper and lower polynomial bounds have been proved, for example, in the case of the fractional
Laplacian perturbed by a negative potential vanishing at infinity, see  \cite[Prop. IV.1 and IV.3]{Carmona-ChenMasters-Simon}
\vspace{0,2cm}

The key observable giving information on the spectral properties of  $\mathrm{H}_{\upalpha,\omega}$ is the fractional average Green's function
$ \mathbb{E}\qty[\abs{\mathrm{G}_z(x_0,x)}^{s}]$ with $0<s<1,$ where 
\[
\mathrm{G}_z= (\mathrm{H}_{\upalpha,\omega}-z)^{-1}
\]
is a well  defined bounded operator for all $z\in \C \setminus \R.$
In this article  we adapt a strategy developed in \cite{schenker2015large} to bound
$ \mathbb{E}\qty[\abs{\mathrm{G}_z(x,y)}^{s}]  $ by the two point function of a self-avoiding walk (SAW) generated by
$D (x,y) =D_{\upalpha,s} (x,y):= | (-\Delta)^{\upalpha } (x,y) |^{s}$. We use this bound to enlarge the set of values for $\lambda $
where pure point spectrum occurs, and to derive improved decay estimates on the corresponding eigenvectors.\vspace{0,2cm}

To formulate our main result we need  some notions  to describe a SAW with long jumps generated by $D.$ These are collected in  Section \ref{sectprelim}.
In particular we denote by % $\chi^{\upalpha,SAW\mathrm{SAW}}$ 
$\chi^{\mathrm{SAW}}_{D}$ the susceptibility,
with radius of convergence %$\mathrm{R}_{\chi^{\upalpha,SAW}}$
$\mathrm{R}_{\chi^{\mathrm{SAW}}_{D}}$, and by
$\mathrm{C}^{\mathrm{SAW}}_{D,\gamma}(x)$
the two-point correlation function with parameter $\gamma>0$,
see \eqref{def:twopf} and \eqref{eq:susceptibility} below. With this notation, our main result is
summarized in the following theorem.

%%%%%%%%%%%%%%%%%%%%%%%%%%%%%%%%%%%%%%%%%%%%%%%%%%%%%%%%%%%%%%%%%%
%%%%%%%%%%%%%%%%%%%%%%%%%%%%%%%%%%%%%%%%%%%%%%%%%%%%%%%%%%%%%%%%%%%%
\begin{theorem}\label{Thm.Schenker}
Assume the one site probability measure  $\mathrm{P}_{0}$ is $\tau-$regular, for some
$\tau\in \left(\frac{d}{d+2\upalpha },1\right)$ with $\tau-$constant $M_{\tau }(\mathrm{P}_0).$
For $s\in \left(\frac{d}{d+2\upalpha },\tau \right)$ we  consider the self-avoiding walk generated by
 $D (x,y)=D_{\upalpha ,s} (x,y):=  |(-\Delta)^{\upalpha } (x,y) |^{s},$ with $x\neq y$.
 Set
\begin{equation}\label{lambda0def}
\theta_{s}:= \frac{\tau }{\tau -s}M_{\tau }(\mathrm{P}_0)^{\frac{s}{\tau }},\qquad \lambda_{0} (s):=\left(\frac{\theta_{s}}{\mathrm{R}_{\chi^{\mathrm{SAW}}_{D}}}\right)^{\frac{1}{s}}.
\end{equation}
%and 
%\begin{equation}\label{lambda0def}
%\lambda_{0} (s):=\left(\frac{\theta_{s}}{\mathrm{R}_{\chi^{\mathrm{SAW}}_{D}}}\right)^{\frac{1}{s}}.
%\end{equation}
Then for  all $\uplambda>\lambda_{0} (s)$  and $  \forall x\neq x_0\in\Z^{d}$ it holds
\begin{equation}\label{Schenker.Bound0}
  \mathbb{E}\qty[\abs{\mathrm{G}_z(x_0,x)}^{s}]
    \leq \ \gamma \ \mathrm{C}^{\mathrm{SAW}}_{D,\gamma}(x-x_0),
\end{equation}
uniformly in $z\in\C\setminus\R,$ where 
\begin{equation}\label{def:gamm}
\gamma=\gamma(\lambda,s):=\frac{\theta_{s}}{\lambda^{s}}. 
\end{equation}

\end{theorem}

%%%%%%%%%%%%%%%%%%%%%%%%%%%%%%%%%%%%%%%%%%%%%%%%%%%%%%%%%%%%%%%%%%%%%%%%%%

%%%%%%%%%%%%%%%%%%%%%%%%%%%%%%%%%%%%%%%%%%%%%%%%%%%%%%%%%%%%%%%%%%
%%%%%%%%%%%%%%%%%%%%%%%%%%%%%%%%%%%%%%%%%%%%%%%%%%%%%%%%%%%%%%%%%%%%
The proof is in Section \ref{sectionSAW}.
Note that  the kernel $D$ is summable by \eqref{eq:deltassummable}. Moreover,
translation invariance of $(-\Delta)^{\upalpha }$ implies that 
$D$ is translation invariant too. The condition $\lambda >\lambda_{0} (s)$ ensures that the two point function is finite
$\mathrm{C}^{\mathrm{SAW}}_{D,\gamma}(x)<\infty $ $\forall x\in \mathbb{Z}^{d}$
(cf. Section \ref{sectprelim} below). The next result uses the bound \eqref{Schenker.Bound0} to prove existence of pure point spectrum and
decay estimates on the eigenvectors.

%%%%%%%%%%%%%%%%%%%%%%%%%%%%%%%%%%%%%%%%%%%%%%%%%%%%%%%%%%%%%%%%%%%%%%%%%%%%%%%%%%%%
%%%%%%%%%%%%%%%%%%%%%%%%%%%%%%%%%%%%%%%%%%%%%%%%%%%%%%%%%%%%%%%%%%%%%%%%%%%%%%%%%%%%%%%%%%%
\begin{theorem}\label{Main.Cor.}
For $s\in \left(\frac{d}{d+2\upalpha },\tau \right)$ we define 
 \begin{align}\label{def.alpha_s}
            2\upalpha_s:=s(d+2\upalpha)-  d,
        \end{align}
which satisfies, by the constraints on $s,$ the inequality 
\[
0<\upalpha_{s}<\upalpha<1. 
\]
Remember the definitions of $\lambda_{0} (s) $  and $\theta_{s}$   \eqref{lambda0def} above.
 For all $\lambda >\lambda_{0} (s)$ the following statements hold.
\begin{itemize}
\item [$(i)$] The averaged fractional Green's function satisfies
\begin{equation}\label{Schenker.Bound}
 \mathbb{E}\qty[\abs{\mathrm{G}_z(x,y)}^{s}]
    \leq  \frac{ K_{0}\theta_{s}}{\lambda^{s}} \ \frac{1}{\abs{x-y}^{d+2\upalpha_{s}}}\qquad
    \forall x,y\in\Z^{d},\: x\neq y,
\end{equation}
uniformly  in $z\in\C\setminus\R.$ The constant $K_{0}=K_{0} (\lambda,s )>0$ is defined in \eqref{K0def} below.

\item [$(ii)$] The spectrum of $\mathrm{H}_{\upalpha,\omega}$ consists only of pure point spectrum  for a.e.
$\omega\in\Upomega.$ 

\item [$(iii)$] Assume the density of $\mathrm{P}_0$ is bounded (in particular $\tau=1$).
Then, for a.e. $\omega\in\Upomega$ and for all  $\mathrm{E}$ in the almost sure spectrum $\Sigma$, there is a localization
 center $\mathrm{x}_{\mathrm{E}}(\omega)\in\Z^{d}$  such that the corresponding normalized eigenfunction $\upvarphi_{\mathrm{E}}(\cdot,\omega)$ satisfies $\forall y\in\Z^{d}$
\begin{align}\label{Ineq.eigenfunction.decay}
   \abs{\upvarphi_{\mathrm{E}}(y,\omega)}^2\leq 4\mathrm{A}_{t}(\omega) \qty[\tfrac{(-\Delta)^{\upalpha}(0,0)}{- (-\Delta)^{\upalpha}(0,\mathrm{x}_{\mathrm{E}}(\omega))}]^2&\frac{1}{\qty(1+\abs{y-\mathrm{x}_{\mathrm{E}}(\omega)})^t},\quad \forall 0<t<2\upalpha_{s},
\end{align}
where $\mathrm{A}_{t}$ is an integrable random variable.

\item [$(iv)$]  Assume the density of $\mathrm{P}_0$ is bounded (in particular $\tau=1$).
For $0<\beta <2\alpha $ we have  for a.e. $\omega\in\Upomega$ 
\begin{equation}\label{eq:moments}
\sup_{t\in \R}\sum_{x\neq 0} |x|^{\beta } \left |e^{-itH_{\upalpha,\omega }} (0,x) \right|^{2} <  \infty.
\end{equation}
In particular, if $\upalpha>\frac{1}{2},$ the first moment of the wave function initially localized at the origin and evolving under the action of
$H_{\upalpha,\omega }$ is finite.

\end{itemize}

\end{theorem}

%%%%%%%%%%%%%%%%%%%%%%%%%%%%%%%%%%%%%%%%%%%%%%%%%%%%%%%%%%%%%%%%%%%%%%%%
\paragraph{Remark.} Whenever the bound in \eqref{eq:moments} holds for all $\beta >0$ the system is said to exhibit
dynamical localization (see \cite{kirsch-panorama} \cite{klein-panorama}).
Here we only prove a weaker version bounding moments for $\beta <2\upalpha.$
In particular, since $\beta <2,$ our bounds do not imply vanishing of the d.c. electrical conductivity (see \cite{aizenman1998localization}).  
Due to the polynomial decay of the fractional Laplacian we do not expect true dynamical localization to hold in this case.

%%%%%%%%%%%%%%%%%%%%%%%%%%%%%%%%%%%%%%%%%%%%%%%%%%%%%%%%%%%%%%%%%%%%%%%
\begin{proof}
$(i)$ follows directly from the bound \eqref{Schenker.Bound0} together
with the decay \eqref{eq:decay2pfunct}.
To prove $(ii)$ and $(iii)$ note that the bound \eqref{Schenker.Bound} ensures that
for all $\lambda >\lambda_{0}(s),$
\begin{equation}\label{decayt}
\sum_{x\in \Z^{d}}\mathbb{E}\qty[\abs{\mathrm{G}_z(0,x)}^{s}|x|^{t}]<\infty \qquad
\forall 0\leq t<2\upalpha_{s}
\end{equation}
holds uniformly in $z \in \C \setminus \R.$ Note that $\lim_{\eta\downarrow 0}\sum_{y\in\Z^{d}}\abs{\mathrm{G}_{\mathrm{E}+i\eta}(x,y)}^2$ always exists (but may be infinite) since
$\sum_{y\in\Z^{d}}\abs{\mathrm{G}_{\mathrm{E}+i\eta}(x,y)}^2=[(E-\mathrm{H}_{\upalpha,\omega})^{2}+\eta^{2}]^{-1} (x,x)$ which is a monotone function in $\eta.$
We argue
\begin{align*}
  &\mathbb{E}\qty[\qty(\lim_{\eta\downarrow 0}\sum_{y\in\Z^{d}}\abs{\mathrm{G}_{\mathrm{E}+i\eta}(x,y)}^2)^{\frac{s}{2}}]=
 \mathbb{E}\qty[\lim_{\eta\downarrow 0}\qty(\sum_{y\in\Z^{d}}\abs{\mathrm{G}_{\mathrm{E}+i\eta}(x,y)}^2)^{\frac{s}{2}}]\\
 &\qquad \leq 
\mathbb{E}\qty[\lim_{\eta\downarrow 0}\qty(\sum_{y\in\Z^{d}}\abs{\mathrm{G}_{\mathrm{E}+i\eta}(x,y)}^{s})]
\leq \liminf_{\eta\downarrow 0}  \sum_{y\in\Z^{d}}
  \mathbb{E}[\abs{\mathrm{G}_{\mathrm{E}+i\eta}(x,y)}^s]<\infty,
\end{align*}
where in the first two steps we used that the function $x\mapsto x^{\frac{s}{2}}$ is monotone and
 $(\sum_{n}a_{n})^{s}\leq \sum_{n}a_{n}^{s}$ for all $a_{n}\geq 0$ and $0<s<1.$ The last two inequalites
follow by Fatou and \eqref{decayt} with $t=0.$
Since we assumed the one site probability measure  $\mathrm{P}_0$ has a density, this bound implies by Simon-Wolff criterion \cite[Thm. 5.7]{aizenman2015random} that the spectrum is pure point only. 
Finally, since  $\mathrm{P}_0$ has compact support and bounded density  we have
\[
\sup_{v\in \R} (1+|v|)^{s} \mathrm{P}_0 (v)<\infty.
\]
The eigenfunction decay follows from this bound and \eqref{decayt}
by standard arguments (cf.  \cite[Thm. 7.4]{aizenman2015random}).
To prove $(iv)$ note that
\[
\sup_{t\in \R}\left |e^{-itH_{\upalpha,\omega }} (0,x) \right|\leq \sup_{F\in C (\R ), \|F\|_{\infty }\leq 1}
 |F (H_{\upalpha,\omega }) (0,x) |=:Q (x,0,\Sigma)
\]
where $Q (x,0,\Sigma)$ is called the eigenfunction correlator (cf.~\cite[eq.~7.1]{aizenman2015random} ) and $\Sigma $ is the almost sure spectrum of $H_{\upalpha,\omega }$.
Using  the bound
\[
\mathbb{E}[ Q (x,0,\Sigma) ]\leq C_{s} \liminf_{\eta\downarrow 0} \int_{\Sigma } 
  \mathbb{E}[\abs{\mathrm{G}_{\mathrm{E}+i\eta}(x,0)}^s]\  d\mathrm{E}%\leq
 % \frac{C_{s} K_{0}\theta_{s}}{\lambda^{s}} \ \frac{1}{\abs{x-y}^{d+2\upalpha_{s}}}
\]
where $C_{s}>0$ is some constant (cf. \cite[Thm. 7.7]{aizenman2015random}), and the decay in $(i)$ we obtain
\[
\sum_{x\neq 0} |x|^{\beta }\mathbb{E}[ Q (x,0,\Sigma) ]\leq  \frac{C_{s} K_{0}\theta_{s}}{\lambda^{s}}
\sum_{x\neq 0} |x|^{\beta }\frac{1}{\abs{x-y}^{d+2\upalpha_{s}}}<\infty 
\]
for all $\beta< s (d+2\upalpha )-d.$ Using $0\leq Q (x,0,\Sigma)\leq 1,$ we conclude
\[
\sup_{t\in \R}\sum_{x\neq 0} |x|^{\beta } \left |e^{-itH_{\upalpha,\omega }} (0,x) \right|^{2} \leq \sum_{x\neq 0} |x|^{\beta } Q_{\omega} (x,0)^{2}\leq  \sum_{x\neq 0} |x|^{\beta } Q_{\omega} (x,0)<\infty 
\]
for a.e. $\omega\in  \Upomega, $ from which the result follows. 
\end{proof}
%%%%%%%%%%%%%%%%%%%%%%%%%%%%%%%%%%%%%%%%%%%%%%%%%%%%

%%%%%%%%%%%%%%%%%%%%%%%%%%%%%%%%%%%%%%%%%%%%%%%%%%%%%%%%%%%%%
\paragraph{Discussion of the results.}
To compare with previous results let us  assume that the probability distribution has bounded density and hence
$\tau=1$ and  $\frac{d}{d+2\upalpha }<s<1.$ In particular $2\upalpha_{s}$ can be arbitrarily close to $2\upalpha.$

A direct application of \cite[Thm.1']{aizenman1998localization} with $|x-y|$ replaced by  $\ln (1+|x-y|)$ and
$K (x,y)$ replaced by $(-\Delta)^{\upalpha }(x,y$) gives the following  estimate $\forall x,x_0\in\Z^{d}$
\begin{equation}\label{AG.Bound}
\mathbb{E}[\abs{\mathrm{G}_z(x_0,x)}^s]\  
            \leq \  M_{1 }(\mathrm{P}_0)^{s} \tfrac{1 }{1 -s}\ 
                \frac{1}{\qty(1+\abs{x-x_0})^{\upbeta}}, \qquad \forall 0<\beta<2\upalpha_{s} 
\end{equation}
which holds uniformly  in $z\in \C \setminus \R, $ as long as $\lambda$ satisfies
\begin{equation}\label{def:lambdaAG}
\lambda >\lambda_{AG} (\beta,s ) :=M_{1 }(\mathrm{P}_0)\left(\frac{2 }{1 -s}
\sum_{x\in \Z^{d}}| (-\Delta )^{\upalpha } (0,x) |^{s} (1+|x|)^{\beta }\right)^{\frac{1}{s}},
\end{equation}
 where the condition $\beta<2\upalpha_{s} $ is needed to ensure that the sum above converges.
This can be seen also in  \cite[Thm. 10.2]{aizenman2015random}: the power $\beta $ here
plays the role of $\mu$ in  \cite[eq. (10.11), (10.12)]{aizenman2015random}.
Note that, while equation (10.12) in   \cite{aizenman2015random} shows that the fractional Green's function decays
algebraically with  power $\beta <2\upalpha_{s},$ the power obtained in Theorem \ref{Main.Cor.}, equation \eqref{Schenker.Bound}, is   $2\upalpha_{s}+d.$
Therefore  we gain  at least a factor $d$ in the decay of the fractional Green's function. 
Since $\tau=1,$ we can get arbitrarily close to the decay 
$(d+2\upalpha ),$  which is  also the decay of the massive resolvent  $((-\Delta )^{\upalpha }+m^{2})^{-1}$  and  of $(-\Delta )^{\upalpha },$ see \eqref{eq:massiveresolv} and \eqref{Eq.decay}.
Moveover, our bound  \eqref{Schenker.Bound} is summable
for all $\upalpha >0,$ which is not the case in \cite{aizenman1998localization}. Indeed  the bound  \eqref{AG.Bound}
is never summable when $\upalpha<\frac{d}{2},$
hence   Simon-Wolff criterion cannot be  applied directly.
One proves instead  the inequality \cite{aizenman1993localization}
\[
\sup_{\eta >0} \mathbb{E}\left[ \sum_{y\in\Z^{d}}
  \abs{\mathrm{G}_{\mathrm{E}+i\eta}(x,y)}^s(1+\abs{x-y})^{\beta }\right ]<\infty,
\]
for $\lambda >\lambda_{AG} (\beta,s ). $
This estimate ensures the existence of pure point spectrum and the decay \eqref{Ineq.eigenfunction.decay}
for the eigenfunctions.

As in \cite{schenker2015large}, the disorder threshold $\lambda_{0} (s)$ we obtain is better than previous results, namely
$\lambda_{AM}$ in \eqref{lambda0def} and $\lambda_{AG}$ in \eqref{def:lambdaAG}. 
Indeed, using \eqref{lambda0def},  \eqref{boundR} and \eqref{lambdaAM} we obtain
\begin{align*}
\lambda_{0} (s) &= M_{1 }(\mathrm{P}_0)\left(\frac{1 }{1 -s} \frac{1}{\mathrm{R}_{\chi^{\mathrm{SAW}}_{D}}}\right)^{\frac{1}{s}}
\leq M_{1 }(\mathrm{P}_0) \left(\frac{1 }{1 -s} \sum_{x\neq 0}  D (0,x)\right)^{\frac{1}{s}}\\
&= M_{1 }(\mathrm{P}_0)\left(\frac{1 }{1 -s} \sum_{x\neq 0}  | (-\Delta )^{\upalpha } (0,x) |^{s} \right)^{\frac{1}{s}} \\
&< M_{1}(\mathrm{P}_0)\left(\frac{2 }{1 -s} \sum_{x\neq 0}  | (-\Delta )^{\upalpha } (0,x) |^{s} \right)^{\frac{1}{s}}=\lambda_{AM} (s)\\
&<M_{1}(\mathrm{P}_0)\left(\frac{2 }{1-s}
\sum_{x\in \Z^{d}}| (-\Delta )^{\upalpha } (0,x) |^{s} (1+|x|)^{\beta }\right)^{\frac{1}{s}}= \lambda_{AG} (s,\beta).
\end{align*}
Hence $\lambda_{0} (s)<\lambda_{AM} (s)<\lambda_{AG} (s,\beta )$ for all $0<\beta< 2\upalpha_{s} $ and $s\in (0,1).$

One may try  to obtain the lowest possible $\lambda_{0}$ by optimizing on $s,$ as in 
 \cite{schenker2015large}. For simplicity we assume the probability density is such that $ M_{1}(\mathrm{P}_0)=1,$
 as  for example when $\mathrm{P}_0$ is the uniform distribution on $[-1,1].$
We need $s\in (\frac{d}{d+2\upalpha },1)$ and $\lambda >0$ to satisfy   $f (s,\lambda )<1,$ where
\[
f (s,\lambda):= \frac{1}{1-s} \frac{1}{\lambda^{s}} \frac{1}{\mathrm{R}_{\chi^{\mathrm{SAW}}_{D_{\upalpha,s }}}}.
\]
In the case $\upalpha =1$ the factor $\mathrm{R}_{\chi^{\mathrm{SAW}}_{D_{\upalpha,s }}}$ is replaced by $\mu_{d}^{-1}$ where
$\mu_{d}$ is the connectivity constant of the simple self-avoiding random walk in $\Z^{d}$ (cf.  \cite{schenker2015large}).
The strategy now is to construct a solution $s_{\lambda }\in C^{1} ((1,\infty );(0,1))$ of the equation $\partial_{s}f (s,\lambda )=0,$ and
then find  $\lambda_{0}$ solving $f (s_{\lambda_{0} },\lambda_{0})=1.$ Since
\[
\frac{d}{d\lambda } f (s_{\lambda },\lambda)=s_{\lambda }' \partial_{s}f (s_{\lambda },\lambda )- \frac{s}{(1-s)} \frac{1}{\lambda^{s+1} }  \frac{1}{\mathrm{R}_{\chi^{\mathrm{SAW}}_{D_{\upalpha,s_{\lambda } }}}}=- \frac{s}{(1-s)} \frac{1}{\lambda^{s+1} }  \frac{1}{\mathrm{R}_{\chi^{\mathrm{SAW}}_{D_{\upalpha,s_{\lambda } }}}}<0 ,
\]
we have $f (s_{\lambda },\lambda )<1$ for all $\lambda>\lambda_{0}.$ For $\upalpha=1,$ since $\mu_{d}$ is independent of
$s,$ it is easy to see that $s_{\lambda}:= 1- (\ln \lambda )^{-1}$ is a solution. However, for $\upalpha<1,$ also the term
$\mathrm{R}_{\chi^{\mathrm{SAW}}_{D_{\upalpha,s }}}$ depends on $s$ in a nontrivial way, which makes this computation very involved.

  We could study instead a function $g (s,\lambda)$ such that  $f \leq g.$
The problem would then be reduced to find $s$ and $\lambda $ such that $g (s,\lambda )<1.$ 
For $d=1,$ using \eqref{boundR} and \eqref{Eq.decay}, a possible choice for $g$  is 
\[
g (s,\lambda )= \frac{1}{1-s}\frac{C_{\upalpha,1 }^{s}}{\lambda^{s} } \left[  1+\frac{2}{(1+2\upalpha )s-1} \right].
\]
Note that the threshold $\lambda_{0}$ we may construct with this procedure will depend on the constant $C_{\upalpha,1 },$
on which we have no information, except in the case $\upalpha=1.$ 

%\newpage

\section{Preliminary definitions and results}\label{sectprelim}
%%%%%%%%%%%%%%%%%%%%%%%%%%%%%%%%%%%%%%%%%%%%%%%%%%%%%%%%%%%%%%%%%%%%%%%%%%%%%%%
\paragraph{$\tau-$regularity and apriori bound.}

%%%%%%%%%%%%%%%%%%%%%%%%%%%%%%%%%%%%%%%%%%%%%%%%%%%%%%%%%%%%%
  \begin{definition}[$\uptau-$regularity]\label{Def.tau.reg.}
    Let $\uptau\in(0,1]$.
    We say that a probability measure $\upmu$ is $\uptau-$regular
    if there is a $\mathbf{\mathrm{C}}>0$
        such that $\upmu\qty([\mathrm{v}-\updelta,\mathrm{v}+\updelta])
                    \leq \mathbf{\mathrm{C}}\updelta^\uptau, \:\forall \mathrm{v}\in\R,\:\forall \updelta>0.$
    If $\upmu$ is $\uptau-$regular, the corresponding $\tau-$constant is defined by
    \begin{align}
        \mathrm{M}_\uptau(\upmu):=\inf\qty{\mathbf{\mathrm{C}}>0\:\vert \:
                                                \upmu\qty([\mathrm{v}-\updelta,\mathrm{v}+\updelta])
                                                \leq \mathbf{\mathrm{C}}\updelta^\uptau,\:\forall \mathrm{v}\in\R,\:\forall \updelta>0}.
    \end{align}
\end{definition}
%%%%%%%%%%%%%%%%%%%%%%%%%%%%%%%%%%%%%%%%%%%%%%%%%%%%%%%%%%%%%%%%%%%%%%
The $\tau-$regularity of $\mathrm{P}_{0}$ enters in the bounds for $G_{z}$ via the so-called \textit{a priori bound}
\begin{equation}\label{boundGxx}
\mathbb{E}\qty[\abs{\mathrm{G}_z(x,x)}^{s}| \omega_{\Z^{d}\setminus \{x \}}]\leq  \frac{\uptheta_{s}}{\lambda^{s}},\qquad \forall x\in \Z^{d},\ 0<s<\tau,
\end{equation}
uniformly in $\omega_{\Z^{d}\setminus \{x \}},$ where $ \uptheta_{s}= \frac{\tau }{\tau -s} M_{\tau } (\mathrm{P}_{0})^{\frac{s}{\tau }}$ (cf. equation \eqref{lambda0def}).
This bound is obtained remarking that $|\mathrm{G}_z(x,x)|= (\lambda |\omega_{x}+\eta (\omega_{\Z^{d}\setminus \{x \}}) |)^{-1}$
where the random complex number $\eta (\omega_{\Z^{d}\setminus \{x \}})$ is independent of $\omega_{x}.$ Since $\omega_{x}$
and $\omega_{\Z^{d}\setminus \{x \}}$ are independent,  the problem reduces to the following estimate
\begin{equation}\label{DecouplingLemma}
 \int_{\R } \frac{1}{\abs{\mathrm{v}-\eta}^s}d\mathrm{P}_{0} (v)\leq \uptheta_{s},\qquad  \forall\eta\in\C
\end{equation}
which holds $\forall 0<s<\tau$ (cf.\cite[App. B]{aizenman1998localization}).

%%%%%%%%%%%%%%%%%%%%%%%%%%%%%%%%%%%%%%%%%%%%%%%%%%%%%%%%%%%%%%%%%%%%%%%%%%%%%%%%
\paragraph{Self-avoiding walks with long-range jumps.}

Let $D\in [0,\infty )^{\Z^{d}\times \Z^{d}}$  be an infinite matrix. Assume $D$ is translation invariant and
\[
0<\sum_{x\neq 0}D (0,x)<\infty.
\]
We consider the  random walk on $\Z^{d}$ with transition probability from $x$ to $y\neq x$
\[
p (x,y)=\frac{D(x,y)}{\sum_{z\neq x}  D(x,z)}.
\]
For 
 $x,x_0\in\Z^{d}$  we consider for $n\geq 1$
\[
\mathcal{W}_n(x_0,x):=\qty{w=\qty(w_j)_{j=0}^n\subset\Z^{nd}\:|\: w_0=x_0,\:w_n=x}
\]
the set of paths in $\Z^{d}$  going from $x_{0}$ to $x$ in $n\geq 1$ steps.
For $x=x_{0}$ we may also have paths of length zero $\mathcal{W}_0(x_0,x_{0}):=\{x_{0} \}.$
We say that $w\in\mathcal{W}_n(x_0,x)$ is a self-avoiding walk (SAW) of length $n$  if $w_k\neq w_l$ for all $k\neq l$ with $k,l\leq n$. The set of self-avoiding paths in $\Z^{d}$  going from $x_{0}$ to $x$ in $n\geq 0$ steps is denoted by $\mathcal{W}_n^{\mathrm{SAW}}(x_0,x).$
Note that $\mathcal{W}_0^{\mathrm{SAW}}(x_0,x_{0})=\mathcal{W}_0(x_0,x_{0})$ and
\[
\mathcal{W}_n^{\mathrm{SAW}}(x_0,x_{0})=\emptyset= \mathcal{W}_0^{\mathrm{SAW}}(x_0,x)\qquad \forall
x_{0}\neq x, n\geq 1.
\]
Following \cite{schenker2015large}, we define, for $n\geq 0,$ and $x\in \Z^{d}$
\begin{align}
          \mathrm{c}^{D}_n(x)=   \mathrm{c}^{\mathrm{SAW}}_{D,n}(x)
                &:=\sum_{w\in\mathcal{W}_n^{\mathrm{SAW}}(0,x)}\prod_{j=0}^{n-1}
                    D(w_{j},w_{j+1}), \label{eq:deftwopoint}
\end{align}
where we took the convention that the sum over an empty set equals $0$ and the product over an empty set equals $1.$
In particular $ \mathrm{c}^{D}_0(x)=\delta_{x,0}$ and $c_{n}^{D} (0)=0$ $\forall n\geq 1.$
The  function $\mathrm{c}^{D}_n(x)$ for $n\geq 1,$
is proportional to the probability that  a self-avoiding random walk
goes from $0$ to $x$ in $n$ steps.
The corresponding  two-point correlation function 
is defined as 
\begin{equation}\label{def:twopf}
             \mathrm{C}^{D}_\gamma(x)=  \mathrm{C}^{\mathrm{SAW}}_{D,\gamma}(x)
            :=\sum_{n\geq 0}\mathrm{c}^{D}_n(x)\gamma^n,
        \quad\forall x\in\Z^{d}.
	    \end{equation}
The sum starts at $n=1$ when $x\neq 0,$ while for $x=0$ we have $\mathrm{C}^{D}_\gamma(0)=\gamma^{0}\mathrm{c}^{D}_0(0)=1.$
For the sake of notation we will often write $\mathrm{C}^{D}_\gamma$ instead of $ \mathrm{C}^{\mathrm{SAW}}_{D,\gamma}.$
  %The corresponding  radius of convergence is denoted by $\mathrm{R}_{\mathrm{C}^{D,SAW}(x)}$.
  Note that in \cite{chen2015critical}, the correlation function is defined in a slightly different way, which can be recovered from our definition performing the
  change of variable
    \[
        \gamma\mapsto \Tilde{\gamma}:=\frac{\gamma}{\sum_{z\neq x}  D(x,z)}.
    \]
 Summing over $x$ we obtain the susceptibility
 \begin{align}\label{eq:susceptibility}
     \chi^{D}_{\gamma}=    \chi^{\mathrm{SAW}}_{D,\gamma}:=\sum_{x\in\Z^{d}} \mathbf{\mathrm{C}}^{D}_\gamma(x)
=\sum_{n\geq 0}\gamma^n\sum_{x\in\Z^{d}}\sum_{w\in\mathcal{W}_n^{\mathrm{SAW}}(0,x)}\prod_{j=0}^{n-1}
                    D(w_{j},w_{j+1}),
    \end{align}
The corresponding  radius of convergence is denoted by  $\mathrm{R}_{\chi^{\mathrm{SAW}}_{D}}.$ 
For $n=2$ we have
\begin{align*}
&\sum_{x\in\Z^{d}}\sum_{w\in\mathcal{W}_2^{\mathrm{SAW}}(0,x)}\prod_{j=0}^{n-1}
  D(w_{j},w_{j+1})= \sum_{x\neq 0} \sum_{x_{1}\neq 0,x} D (0,x_{1})D (x_{1},x)\\
&=
 \sum_{x_{1}\neq 0}  D (0,x_{1})     \sum_{x\neq x_{1},0}D (x_{1},x)
   \leq  \sum_{x_{1}\neq 0}  D (0,x_{1})     \sum_{x\neq x_{1}}D (x_{1},x)=
		 \left (\sum_{z\neq 0}  D(0,z) \right)^{2}.
\end{align*}
Repeating this argument for general $n\geq 1$ we obtain 
\[
\sum_{x\in\Z^{d}}\sum_{w\in\mathcal{W}_n^{\mathrm{SAW}}(0,x)}\prod_{j=0}^{n-1}
                    D(w_{j},w_{j+1})
		    \leq  \left (\sum_{z\neq 0}  D(0,z) \right)^{n},
\]
and hence
\begin{equation}\label{boundR}
\mathrm{R}_{\chi^{\mathrm{SAW}}_{D}}\geq \frac{1}{\sum_{z\neq 0}  D(0,z)}>0.
\end{equation}
The decay of the two point function of the SAW introduced above has been estimated in  
 \cite[Lemma 2.4]{chen2015critical}.
 We recall the result here, together with a sketch of the proof, translated into our language.

%%%%%%%%%%%%%%%%%%%%%%%%%%%%%%%%%%%%%%%%%%%%%%%%%
 \begin{lemma}[Decay of the SAW  two point function]\label{le:decay2pfunct} Assume that $D (0,x)\leq \mathcal{C}\frac{1}{|x|^{d+a}}$  holds for some $\mathcal{C},a>0$
 and for all $x\neq 0.$
 Then the  two point function of the SAW generated by $D$
 is bounded by 
\begin{equation}\label{eq:decay2pfunct}
\mathrm{C}^{\mathrm{SAW}}_{D,\gamma}(x)\leq K_{0}\ \frac{1}{\abs{x}^{d+a}},
\end{equation}
for all $\gamma <\mathrm{R}_{\chi^{\mathrm{SAW}}_{D}}.$ The constant $K_{0}=K_{0} (d,a)>0$
can be explicitely written in terms of the susceptibility $\chi^{D}_{\gamma}$ as follows:
\begin{equation}\label{K0def}
K_{0}=\tilde{\ell}^{d+a }  \chi^{D}_{\gamma}+2 (\chi^{D}_{\gamma})^{2}\gamma  \mathcal{C}
\end{equation}
where $\tilde{\ell}=\tilde{\ell} (d,a)>0$ is the minimal distance such that 
\[
c (x):=\sum_{\abs{u}\leq \frac{\abs{x}}{3}\leq \abs{v}} \mathrm{C}^{D}_{\gamma}(u)\ \gamma D(u,v)\leq \frac{1}{2}2^{- (d+a) }\qquad \forall  |x|\geq  \tilde{\ell}.
\]
\end{lemma}
%%%%%%%%%%%%%%%%%%%%%%%%%%%%%%%%%%%%%%%%%%%%%%%%%%%%%
Note that $\tilde{\ell}$ is well defined since  $\lim_{|x|\to \infty }c (x)=0$ (see below).

%%%%%%%%%%%%%%%%%%%%%%%%%%%%%%%%%%%%%%%%%%%%%%%%%%%%%%%
\begin{proof} 
The assumption $\gamma <\mathrm{R}_{\chi^{\mathrm{SAW}}_{D}}$ ensures  $\mathrm{C}^{D}_{\gamma}(x)<\infty $ $\forall x\in \Z^{d}.$ The key ingredient of the proof is the following inequality, which holds for any $0<\ell <|x|$
\begin{equation}\label{keyineq}
  \mathrm{C}^{D}_{\gamma}(x)
        \leq \sum_{\substack{u,v\in\Z^{d}\\\abs{u}\leq \ell<\abs{v}}}
            \mathrm{C}^{D}_{\gamma}(u)\ \gamma D(u,v)\ \mathrm{C}^{D}_{\gamma}(x-v)
\end{equation}
To prove it, remember that 
\[
 \mathrm{C}^{D}_{\gamma}(x)= \sum_{n\geq 0}
 \sum_{w\in\mathcal{W}^{\mathrm{SAW}}_n(0,x)}\gamma^{n}
 \prod_{j=0}^{n-1} D(w_{j},w_{j+1}).
\]
For a given path $w\in\mathcal{W}_n^{SAW}(0,x)$ we define $u:= w_{j_{m}}$ and $v:= w_{j_{m}+1}$
where 
\[
j_m:= \max\{ j\in \{0,\dotsc ,n\}| \ |w_{j}|<\ell\}.
\]
Since $0<\ell <|x|$ this set is non-empty and $0\leq j_{m}<n.$ With this definitions the sum
above can be reorganized as
\begin{align*}
 &\mathrm{C}^{D}_{\gamma}(x)=
 \sum_{\substack{u,v\in\Z^{d}\\ \abs{u}\leq \ell<\abs{v}}}\sum_{n,m\geq 0}
 \sum_{w\in\mathcal{W}^{\mathrm{SAW}}_n(0,u)} \sum_{w'\in\mathcal{W}^{\mathrm{SAW}}_m(v,x)} [\gamma^{n}\prod_{j=0}^{n-1}D(w_{j},w_{j+1})] \gamma D(u,v) \ \cdot \\
&\qquad \ [\gamma^{m}
\prod_{j=0}^{m-1}D(w_{j}',w_{j+1}')]\  
\Ind_{\qty{w\cup w'\textrm{ is SAW}}}\ \leq\
\sum_{\substack{u,v\in\Z^{d}\\\abs{u}\leq \ell<\abs{v}}}
            \mathrm{C}^{D}_{\gamma}(u)\ \gamma D(u,v)\ \mathrm{C}^{D}_{\gamma}(x-v),
\end{align*}
where in the last step we applied $\Ind_{\qty{w\cup w'\textrm{ is SAW}}}\leq 1$ and the translation invariance of $D.$
Set now $\ell=\frac{\abs{x}}{3}$. The sum on the RHS of  \eqref{keyineq} can be reorganized
as follows
\[
\sum_{ \abs{u}\leq \frac{\abs{x}}{3}<\abs{v}}=
 \sum_{\abs{u}\leq \frac{\abs{x}}{3},\:\: \frac{\abs{x}}{2}<\abs{v}} +\sum_{\abs{u}\leq \frac{\abs{x}}{3}< \abs{v}\leq\frac{\abs{x}}{2}} .
\]
We estimate the first sum as follows:
\begin{equation}
  \sum_{\abs{u}\leq \frac{\abs{x}}{3},\: \frac{\abs{x}}{2}<\abs{v}} 
    \mathrm{C}^{D}_{\gamma}(u)\ \gamma D(u,v)\ \mathrm{C}^{D}_{\gamma}(x-v)\leq
   (\chi^{D}_{\gamma})^{2}  \gamma \mathcal{C}\frac{1}{\abs{x}^{d+a}} 
    \end{equation}
where we used \eqref{eq:susceptibility}, $|u-v|\geq |x|/6$ and $D (u,v)\leq\frac{ \mathcal{C}}{|u-v|^{d+a}}.$
The second sum is bounded by
\[
   \sum_{\abs{u}\leq \frac{\abs{x}}{3}< \abs{v}\leq\frac{\abs{x}}{2}} 
\mathrm{C}^{D}_{\gamma}(u)\ \gamma D (u,v)\ \mathrm{C}^{D}_{\gamma}(x-v)\leq
 c (x) \sup_{|v|\leq \frac{\abs{x}}{2}}
    \mathrm{C}^{D}_{\gamma}(x-v)=
 c (x) \sup_{|v|\geq \frac{\abs{x}}{2}}
    \mathrm{C}^{D}_{\gamma}(v),   
\]
where we defined $c (x):=\sum_{\abs{u}\leq \frac{\abs{x}}{3}< \abs{v}} \mathrm{C}^{D}_{\gamma}(u)\ \gamma D(u,v).$ Putting all this together we obtain 
\begin{equation}\label{eq:basicineq}
\mathrm{C}^{D}_{\gamma}(x)\leq   c (x) \sup_{|v|\geq \frac{\abs{x}}{2}}
    \mathrm{C}^{D}_{\gamma}(v)+   (\chi^{D}_{\gamma})^{2}  \gamma \mathcal{C}\frac{1}{\abs{x}^{d+a}} .
\end{equation}
Using the spatial decay of $D$ 
we argue
\begin{align*}
&c (x)\leq  \sum_{\abs{u}<  \frac{\abs{x}}{5},\ \frac{\abs{x}}{3}< \abs{v}} \mathrm{C}^{D}_{\gamma}(u)\ \gamma D(u,v)
+\sum_{\frac{\abs{x}}{5}\leq \abs{u}\leq \frac{\abs{x}}{3}< \abs{v}} \mathrm{C}^{D}_{\gamma}(u)\ \gamma D(u,v)\\
&
\leq \chi^{D}_{\gamma}\sum_{ \abs{v}>\frac{\abs{x}}{5}} \gamma \mathcal{C}\frac{1}{\abs{v}^{d+a}}  +
\sum_{|u|\geq \frac{\abs{x}}{5}}   \mathrm{C}^{D}_{\gamma}(u)
\sum_{|v|\geq 1}   \gamma \mathcal{C}\frac{1}{\abs{v}^{d+a}} \leq
\mathcal{C}' \left[  \frac{1}{ \abs{x}^{a}} +\sum_{|u|\geq \frac{\abs{x}}{5}}   \mathrm{C}^{D}_{\gamma}(u) \right],
\end{align*}
for some constant $\mathcal{C}'.$
Since  $\lim_{\abs{x}\to \infty }\sum_{|u|\geq   \abs{x}}   \mathrm{C}^{D}_{\gamma}(u)=0$ we obtain 
 $\lim_{|x|\to \infty }c (x)=0.$  Hence there is a $\tilde{\ell}=\tilde{\ell} (d,a)$ such that 
\[
c (x)\leq \frac{1}{2}2^{- ( d+a) }\qquad \forall  |x|\geq  \tilde{\ell}.
\]
For $2^{n-1}\tilde{\ell}\leq |x|<2^{n}\tilde{\ell}$ with $n\geq 1 $ we apply $n$ times the inequality
\eqref{eq:basicineq} and obtain
\[
\mathrm{C}^{D}_{\gamma}(x)\leq \frac{1}{2^{n (d+a +1)}}
\sup_{|v|\geq \frac{\abs{x}}{2}} \mathrm{C}^{D}_{\gamma}(v)
+ \Big (\sum_{j=0}^{n}\frac{1}{2^{j}} \Big)\frac{ (\chi^{D}_{\gamma})^{2}\gamma \mathcal{C}}{|x|^{d+a}}
\leq \frac{\tilde{\ell}^{d+a}  \chi^{D}_{\gamma}+2 (\chi^{D}_{\gamma})^{2} \gamma  \mathcal{C}}{|x|^{d+a }}.
\]
When $|x|<\tilde{\ell}$ we apply the simple bound
$\mathrm{C}^{D}_{\gamma}(x)\leq  \chi^{D}_{\gamma}\leq \tilde{\ell}^{d+a }  \chi^{D}_{\gamma}/|x|^{d+a }.$
\end{proof}

%%%%%%%%%%%%%%%%%%%%%%%%%%%%%%%%%%%%%%%%%%%%%%%%%%%%%%%%%%%%%%%%%%%%%%%%%%%%%%%%
\section{Comparison with a long range SAW.}\label{sectionSAW}

%%%%%%%%%%%%%%%%%%%%%%%%%%%%%%%%%%%%%%%%%%%%%%%%%%%%%%%%%%%%%%%%%%%%%%%%%%%%%%%%%%%%%%%

The  proof of Theorem \ref{Thm.Schenker} adapts the strategy of \cite[Thm. 1]{schenker2015large} to the fractional Anderson model. In particular, this requires to work with Green's functions defined  on a volume where at different stages some sites are removed (see also \cite{ASFH}, \cite{hundertmark} \cite{tautenhan}) Therefore, 
for any $\Lambda \subset \Z^{d}$ subset of $\Z^{d}$ (finite or infinite) we introduce the restricted Green's function
$G^{\Lambda }$
\begin{align}\label{Def.G.Box}
  \mathrm{G}^{\Lambda}_{z}(x,y):=
  \begin{cases}
     (\mathrm{H}^{\Lambda}_{\upalpha,\omega  }-z)^{-1}(x,y),&\qquad\forall x,y\in\Lambda,\\
    0,&\qquad\textrm{otherwise.}
  \end{cases}
\end{align}
where $\mathrm{H}^{\Lambda}_{\upalpha,\omega  }-z\in \C^{\Lambda \times \Lambda }$ is the matrix (finite or infinite) obtained by restricting
$\mathrm{H}_{\upalpha,\omega  }-z$ to $\Lambda.$ This matrix is invertible for all $z\in\C\setminus\R.$  
In particular $\mathrm{G}_z^{\Z^{d}}=\mathrm{G}_{z}.$ In the following $\Lambda=\Z^{d},$ but we leave the notation $\Lambda $
through the proof below to stress the fact that the same result holds for any volume.

To simplify the notation  we also set\footnote{Note that this definition differs from the corresponding operator defined via functional calculus by a phase. } 
\[
\fl^\upalpha(x_0,x):=-(-\Delta)^{\upalpha}(x_0,x),\qquad \forall  x,x_0\in\Z^{d}.
\]
Note that  with this convention $\fl^{\upalpha } (x,y)>0$ $\forall x\neq y.$\vspace{0,2cm}

%%%%%%%%%%%%%%%%%%%%%%%%%%%%%%%%%%%%%%%%%%%%%%%%%%%%%%%%%%%%%%%%%%%%%%%%%%%%%%%%%%%%
\begin{proof}[Proof of Theorem \ref{Thm.Schenker}]
By the resolvent identity we have, for all $x\neq x_{0}\in \Lambda $
\begin{align*}
\mathrm{G}_z^{\Lambda}(x_0,x)&= \mathrm{G}_z^{\Lambda}(x_0,x_{0}) \fl^{\upalpha } (x_{0},x)
\mathrm{G}_z^{\Lambda\setminus\qty{x_0}}(x,x)\\
&\quad 
+ \mathrm{G}_z^{\Lambda}(x_0,x_{0}) \hspace{-0,3cm}\sum_{\substack{ w_1\in\Lambda\setminus \{x_0,x\} }}\fl^{\upalpha } (x_{0},w_{1})
\mathrm{G}_z^{\Lambda \setminus\qty{x_0}}(w_{1},x).
\end{align*}
Repeating the procedure $N$ times we obtain
\begin{align*}
&\mathrm{G}_z^{\Lambda}(x_0,x)= \mathrm{G}_z^{\Lambda}(x_0,x_{0})\Big [\sum_{n=1}^{N} \sum_{w\in\mathcal{W}_n^{\mathrm{SAW}}(x_{0},x)}
\prod_{j=0}^{n-1}    \fl^{\upalpha}(w_{j},w_{j+1}) \prod_{j=1}^{n} \mathrm{G}_z^{\Lambda\setminus \{w_{0},\dotsc ,w_{j-1} \}}(w_{j},w_{j})
                 \\
&\quad 
+\hspace{-0,3cm} \sum_{w\in\mathcal{W}_{N+1}^{\mathrm{SAW}}(x_{0},x)}\prod_{j=0}^{N-1}    \fl^{\upalpha}(w_{j},w_{j+1}) 
\prod_{j=1}^{N-1} \mathrm{G}_z^{\Lambda\setminus \{w_{0},\dotsc ,w_{j-1} \}}(w_{j},w_{j})
\mathrm{G}_z^{\Lambda\setminus \{w_{0},\dotsc ,w_{N} \}}(w_{N},x) )\Big ]
\end{align*}
Taking the average and using  the concavity of the function   $y\mapsto y^s$ we have
\begin{align*}
&\mathbb{E}\left[ |\mathrm{G}_z^{\Lambda}(x_0,x)|^{s}\right]\leq \sum_{n=1}^{N} \sum_{w\in\mathcal{W}_n^{\mathrm{SAW}}(x_{0},x)}
\prod_{j=0}^{n-1}    \fl^{\upalpha}(w_{j},w_{j+1})^{s}\cdot  \\
&\cdot 
\mathbb{E}\left[ |\mathrm{G}_z^{\Lambda}(x_0,x_{0})|^{s} \prod_{j=1}^{n}
|\mathrm{G}_z^{\Lambda\setminus \{w_{0},\dotsc ,w_{j-1} \}}(w_{j},w_{j})|^{s}       \right] 
 +\hspace{-0,5cm}\sum_{w\in\mathcal{W}_{N+1}^{\mathrm{SAW}}(x_{0},x)}\prod_{j=0}^{N-1}    \fl^{\upalpha}(w_{j},w_{j+1})^{s} \cdot   \\
&\cdot 
 \mathbb{E}\left[ |\mathrm{G}_z^{\Lambda}(x_0,x_{0})|^{s} 
\prod_{j=1}^{N-1}| \mathrm{G}_z^{\Lambda\setminus \{w_{0},\dotsc ,w_{j-1} \}}(w_{j},w_{j})|^{s}
\ |\mathrm{G}_z^{\Lambda\setminus \{w_{0},\dotsc ,w_{N} \}}(w_{N},x) )|^{s}\right].
\end{align*}
The resolvent $\mathrm{G}_z^{\Lambda\setminus \{w_{0},\dotsc ,w_{j-1} \}}(y,y')$ does not depend on the random variables
$\omega_{w_{i}}, i=0,\dotsc j-1,$ hence recursive applications of the apriori bound \eqref{boundGxx}, which holds since we assume $s< \tau, $ yield
\begin{align*}
&\mathbb{E}\left[ |\mathrm{G}_z^{\Lambda}(x_0,x_{0})|^{s} \prod_{j=1}^{n}
|\mathrm{G}_z^{\Lambda\setminus \{w_{0},\dotsc ,w_{j-1} \}}(w_{j},w_{j})|^{s}       \right] \leq  \left(\frac{\theta_{s}}{\lambda^{s}} \right)^{n+1}\\
&  \mathbb{E}\left[ |\mathrm{G}_z^{\Lambda}(x_0,x_{0})|^{s} 
\prod_{j=1}^{N-1}| \mathrm{G}_z^{\Lambda\setminus \{w_{0},\dotsc ,w_{j-1} \}}(w_{j},w_{j})|^{s}
\ |\mathrm{G}_z^{\Lambda\setminus \{w_{0},\dotsc ,w_{N} \}}(w_{N},x) )|^{s}\right]\leq \left(\frac{\theta_{s}}{\lambda^{s}} \right)^{N}\frac{1}{|\Re z|^{s}}
\end{align*}
where we also applied the inequality $|\mathrm{G}_z^{\Lambda\setminus \{w_{0},\dotsc ,w_{N} \}}(w_{N},x)|\leq\frac{1}{|\Re z|}. $ 
Inserting these estimates in the sums above and using the translation invariance of $\fl^{\upalpha }$ we get
\begin{align*}
\mathbb{E}\left[ |\mathrm{G}_z^{\Lambda}(x_0,x)|^{s}\right]&\leq \frac{\theta_{s}}{\lambda^{s}}
\sum_{n=1}^{N}  \left(\frac{\theta_{s}}{\lambda^{s}} \right)^{n} \sum_{w\in\mathcal{W}_n^{\mathrm{SAW}}(0,x-x_{0})}
\prod_{j=0}^{n-1}  \fl^{\upalpha}(w_{j},w_{j+1})^{s}  \\ 
& +\left(\frac{\theta_{s}}{\lambda^{s}} \right)^{N}\frac{1}{|\Re z|^{s}}\sum_{w\in\mathcal{W}_{N+1}^{\mathrm{SAW}}(0,x-x_{0})}
\prod_{j=0}^{N-1}    \fl^{\upalpha}(w_{j},w_{j+1})^{s}\\
&\leq \frac{\theta_{s}}{\lambda^{s}} C_{\gamma }^{D} (x-x_{0}) +\frac{1}{|\Re z|^{s}} Err (N)
\end{align*}
where $ C_{\gamma }^{D} (x-x_{0})$ is the two point function of the SAW generated by  $D (x,y):= \fl^{\upalpha } (x,y)^{s}=   | (-\Delta)^{\upalpha}(x,y)|^{s}$ with
$\gamma=  \frac{\theta_{s}}{\lambda^{s}}.$ The error term $Err (N)$ satisfies 
\begin{align*}
 Err (N)&:=\left(\frac{\theta_{s}}{\lambda^{s}} \right)^{N} \hspace{-0,3cm}\sum_{w\in\mathcal{W}_{N+1}^{\mathrm{SAW}}(0,x-x_{0})}
\prod_{j=0}^{N-1}    \fl^{\upalpha}(w_{j},w_{j+1})^{s}\\
&\leq
\left(\frac{\theta_{s}}{\lambda^{s}} \right)^{N}\sum_{y\in Z^{d}} \sum_{w\in\mathcal{W}_{N}^{\mathrm{SAW}}(0,y)}
\prod_{j=0}^{N-1}    \fl^{\upalpha}(w_{j},w_{j+1})^{s}.
\end{align*}
Up to now, the above sums may be infinite. The generating kernel for the SAW satisfies
\[
\sum_{y\neq 0} D (y)= \sum_{y\neq 0} \fl^{\upalpha } (0,y)^{s}=\sum_{y\neq 0} |(-\Delta)^{\upalpha } (0,y)|^{s}\leq C  \sum_{y\neq 0} \frac{1}{|x-y|^{(d+2\upalpha )s}}.
\]
The sum above is finite for all $s>\frac{d}{d+2\upalpha }.$ Finally, the assumption $\lambda >\lambda_{0}$ ensures 
that $\gamma <\mathrm{R}_{\chi^{\mathrm{SAW}}_{D}}$ holds and hence $\lim_{N\to \infty }Err (N)=0$ since the susceptibility is finite.
This completes the proof of the theorem.
\end{proof}

%%%%%%%%%%%%%%%%%%%%%%%%%%%%%%%%%%%%%%%%%%%%%%%%%%%%%%%%%%%%%%%%%%%%%%%
%%%%%%%%%%%%%%%%%%%%%%%%%%%%%%%%%%%%%%%%%%%%%%%%%%%%%%%%%%%%%%%%%%%%%%%
%\begin{appendices}
\section{Properties of  the fractional Laplacian.}
\label{sect:fractionalLapl}
%%%%%%%%%%%%%%%%%%%%%%%%%%%%%%%%%%%%%%%%%%%%%%%%%%%%%%%%%%%%%%%%

\subsection{Matrix elements}

In this section we collect some properties of the fractional Laplacian.
The matrix elements of $(-\Delta)^{\upalpha }$ admit the following explicit
representation for all $ x,y\in\Z^{d}$
\begin{equation}\label{Eq.SemigroupFL}
    (-\Delta)^{\upalpha}\qty(x,y)=\frac{-1}{\abs{\Upgamma (-\upalpha)}}\int_0^\infty \frac{\dd t}{t^{1+\upalpha}}
\left[
e^{-2dt}\prod_{j=1}^{d}\mathrm{I}_{\qty(x_j-{y}_j)}\:(2t)-\delta_{xy} \right]
\end{equation}
where $\mathrm{I}_{\mathrm{p}}$ is the modified Bessel function of order $p\in\Z,$ which is defined as 
\begin{align}
    \mathrm{I}_{\mathrm{p}}(t):=\sum_{\mathrm{q}\geq 0}\frac{1}{\mathrm{q}!\Upgamma(\mathrm{p}+\mathrm{q}+1)}\qty(\frac{t}{2})^{2\mathrm{q}+\mathrm{p}}.
\end{align}
This follows from the representation
\begin{align}\label{Bochner.FL}
    (-\Delta)^{\upalpha}=\frac{-1}{\abs{\Upgamma (-\upalpha)}}\int_0^\infty \frac{\dd t}{t^{1+\upalpha}}(e^{t\Delta}-\textrm{Id}),
\end{align}
where the integral converges under the operator norm,
see \cite[Theorem 1.1 (c)]{kwasnicki2017ten},
together with the  relation  (cf. equations (5.3)-(5.4) in \cite{gebert2020lipschitz})
\[
  e^{t\Delta}(x,y)=e^{-2dt}\prod_{j=1}^{d}\mathrm{I}_{\qty(x_j-{y}_j)}\:(2t)\quad \forall x,y\in\Z^{d}.
\]
Note that, since $\Gamma$ has simple poles in the set of the non-positive integers, the equality
\[
\mathrm{I}_{\mathrm{p}}(2t)= \sum_{\mathrm{q}\geq 0, p+q\geq 0}
\frac{1}{\mathrm{q}!(\mathrm{p}+\mathrm{q})!}t^{2\mathrm{q}+\mathrm{p}}
\]
holds for all $p\in \Z .$
In particular this implies  $I_{p} (2t)>0$ $\forall t>0$ and $p\in \Z $ and 
\begin{equation}\label{Ieqsum}
\sum_{p\in \Z} \mathrm{I}_{\mathrm{p}}(2t)= e^{2t}.
\end{equation}
Note that $I_{p}=I_{-p}$ $\forall p>0,$ hence, using also $\Upgamma (n+1)=n!$ for all
$n\geq 0$ we obtain
\begin{equation}\label{Ieqdef}
 \mathrm{I}_{\mathrm{p}}(t)=\sum_{\mathrm{q}\geq 0}
 \frac{1}{\mathrm{q}!(|\mathrm{p}|+\mathrm{q})!}\qty(\frac{t}{2})^{2\mathrm{q}+|\mathrm{p}|}\quad \forall p\in \Z.
\end{equation}

%%%%%%%%%%%%%%%%%%%%%%%%%%%%%%%%%%%%%
\begin{proposition} The matrix elements of $(-\Delta)^{\upalpha }$ satisfy
 $(-\Delta)^{\upalpha } (x,y)<0$ for all $x\neq y$ and 
\begin{align}\label{MeanValueFL}
     (-\Delta)^{\upalpha}(x,x)=-\sum_{y\in\Z^{d}\setminus\qty{x}}(-\Delta)^\upalpha\qty(x,y),
     \qquad \forall x\in\Z^{d}.
\end{align}
In particular $(-\Delta)^{\upalpha } (x,x)= (-\Delta)^{\upalpha } (0,0)>0.$
\end{proposition}
%%%%%%%%%%%%%%%%%%%%%%%%%%%%%%%%%%%%

\begin{proof}
The first statement follows from the fact that $I_{p} (t)>0$  holds $\forall t>0,$ $p\in \Z.$
To prove \eqref{MeanValueFL} we argue
\begin{align*}
-\sum_{y\in\Z^{d}\setminus\qty{x}}(-\Delta)^\upalpha\qty(x,y)&=\frac{1}{\abs{\Upgamma (-\upalpha)}}
\sum_{y\in\Z^{d}\setminus\qty{x}} \int_0^\infty \frac{\dd t}{t^{1+\upalpha}}
e^{-2dt}\prod_{j=1}^{d}\mathrm{I}_{\qty(x_j-{y}_j)}\:(2t)\\
&= \frac{1}{\abs{\Upgamma (-\upalpha)}}\int_0^\infty \frac{\dd t}{t^{1+\upalpha}}
e^{-2dt}\sum_{y\in\Z^{d}\setminus\qty{x}}\prod_{j=1}^{d}\mathrm{I}_{\qty(x_j-{y}_j)}\:(2t).
\end{align*}
Using  $I_{p}=I_{-p}$ and \eqref{Ieqsum} we compute
\begin{align*}
\sum_{y\in\Z^{d}\setminus\qty{x}}\prod_{j=1}^{d}\mathrm{I}_{\qty(x_j-{y}_j)}(2t)&=
\sum_{y\neq 0}\prod_{j=1}^{d}\mathrm{I}_{{y}_j}(2t)\\
&= \sum_{y\in\Z^{d}}\prod_{j=1}^{d}\mathrm{I}_{{y}_j}(2t)- \mathrm{I}_{0}(2t)^{d}=
e^{2dt}-\mathrm{I}_{0}(2t)^{d},
\end{align*}
and hence
\[
-\sum_{y\in\Z^{d}\setminus\qty{x}}(-\Delta)^\upalpha\qty(x,y)=
 \frac{1}{\abs{\Upgamma (-\upalpha)}}\int_0^\infty \frac{\dd t}{t^{1+\upalpha}}
 \left[1- e^{-2dt}\mathrm{I}_{0}(2t)^{d} \right]= (-\Delta)^\upalpha\qty(x,x).
\]
This concludes the proof.
\end{proof}
The limits $\upalpha \to 0$ and $\upalpha\to 1$ can be controlled. This is the content of the next proposition. The proof extends the strategy of \cite[Thm.  1.2]{ciaurri2015fractional} to the case $d\in\N$.

%%%%%%%%%%%%%%%%%%%%%%%%%%%%%%%%%%555
%%%%%%%%%%%%%%%%%%%%%%%%%%%%%%%%%%%%%%%%%%%%%%%%%%%%%%%%%%%%%%5
\begin{theorem}
The matrix elements of $(-\Delta)^{\upalpha }$ satify,\
\begin{align}\label{bound1}
\lim_{\upalpha \to 1}\sup_{|x|>1}|(-\Delta)^{\upalpha}(x,0)|=0,\quad \lim_{\upalpha \to 0}\sup_{|x|>1}|(-\Delta)^{\upalpha}(x,0)|=0&\qquad
\mbox{if }  |x|>1\\
 \lim_{\upalpha \to 1}|(-\Delta)^{\upalpha}(x,0)|=1,\quad \lim_{\upalpha \to 0}|(-\Delta)^{\upalpha}(x,0)|=0&\qquad
\mbox{if }  |x|=1\label{bound2}\\
\lim_{\upalpha \to 1}|(-\Delta)^{\upalpha}(0,0)|=2d,\quad \lim_{\upalpha \to 0}|(-\Delta)^{\upalpha}(0,0)|=1& \qquad
\mbox{if }  |x|=0.\label{bound3}
\end{align}

\end{theorem}
%%%%%%%%%%%%%%%%%%%%%%%%%%%%%%%%%%%%%%%%%%%%%%%%%%%%%%
Note that this result implies, in particular, using \eqref{MeanValueFL},
\[
\lim_{\upalpha \to 1}\sum_{x\neq 0} |(-\Delta)^{\upalpha}(x,0)|=2d, \quad \lim_{\upalpha \to 0}\sum_{x\neq 0} |(-\Delta)^{\upalpha}(x,0)|=1.
\]
\begin{proof}
Remember that  for $x= (x_{1},\dotsc ,x_{d})$ we defined
$|x|=|x|_{2}= (\sum_{j=1}^{d}x_{j}^{2})^{\frac{1}{2}}.$
Set also $|x|_{1}:= \sum_{j=1}^{d}|x_{j}|.$

Using \eqref{Eq.SemigroupFL} 
and \eqref{Ieqdef}, we have
\begin{align*}
&\abs{\Upgamma (-\upalpha)}\, |(-\Delta)^{\upalpha}(x,0)|=
\int_0^\infty \frac{\dd t}{t^{1+\upalpha}}\left[e^{-2dt}
\prod_{j=1}^{d}\mathrm{I}_{x_{j}}(2t)-\delta_{0x} \right]\\
&= \int_0^\infty \frac{\dd t}{t^{1+\upalpha}}\left[e^{-2dt}
\prod_{i=1}^{d}\frac{1}{|x_{i}|!}  \ t^{|x|_{1}} -\delta_{0x} \right]
+
\sum_{|q|_{1}\geq 1}
\prod_{i=1}^{d}\frac{1}{q_{i}!(q_{i}+|x_{i}|)!} \int_0^\infty 
\dd t  \ t^{2|q|_{1}+|x|_{1}-1-\upalpha}e^{-2dt}\\
&= S_{1} (\upalpha, |x|_{1})+ S_{2} (\upalpha, |x|_{1}),
\end{align*}
where
\begin{align}
S_{1} (\upalpha, 0)=&  \int_0^\infty 
\dd t  \ t^{-1-\upalpha}\left( 1- e^{-2dt} \right),   \label{defS10}\\
S_{1} (\upalpha, |x|_{1})=&\prod_{i=1}^{d}\frac{1}{|x_{i}|!}\ \frac{1}{(2d)^{|x|_{1}-\upalpha}}
\Upgamma (|x|_{1}-\upalpha  )\quad \mbox{for } |x|_{1}>0, \label{defS1}
\end{align}
and for all $|x|_{1}\geq 0$
\begin{equation}\label{defS2}
S_{2} (\upalpha, |x|_{1})= \sum_{|q|_{1}\geq 1}
\prod_{i=1}^{d}\frac{1}{q_{i}!(q_{i}+|x_{i}|)!}\ \frac{1}{(2d)^{2|q|_{1}+|x|_{1}-\upalpha}}
\Upgamma (2|q|_{1}+|x|_{1}-\upalpha  ).
\end{equation}
We claim that, $\forall |x|_{1}\geq 0,$
\begin{equation}\label{eq:claimS2}
0\leq \frac{S_{2} (\upalpha, |x|_{1})}{\abs{\Upgamma (-\upalpha)}  }\leq
\frac{ (2d)^{\upalpha } (1-\upalpha )  \Upgamma (1-\upalpha  ) }{\abs{\Upgamma (-\upalpha)}  } C'_{d}= (2d)^{\upalpha } \upalpha  (1-\upalpha  ) C'_{d},
\end{equation}
for some constant $C'_{d}>0$  independent of $\upalpha $ and $x.$ This implies
\begin{equation}\label{S2bound}
\lim_{\upalpha \to 0} \frac{S_{2} (\upalpha, |x|_{1})}{\abs{\Upgamma (-\upalpha)}  }=0=\lim_{\upalpha \to 1}
\frac{S_{2} (\upalpha, |x|_{1})}{\abs{\Upgamma (-\upalpha)} }\qquad  \forall |x|_{1}\geq 0.
\end{equation}
\vspace{0,2cm}

To prove the claim, note that, since $|q|_{1}\geq 1$ we have $2|q|_{1}+|x|_{1}\geq 2$ and hence,
using  
\[
\Upgamma (z+n)=z (z+1)\dotsb (z+n-1)\Upgamma (z)= z\Upgamma (z) \prod_{l=1}^{n-1} (z+l),
\]
and $0< \upalpha <1,$ we get
\begin{align*}
&\Upgamma (2|q|_{1}+|x|_{1}-\upalpha  ) =   (1-\upalpha ) \Upgamma (1-\upalpha  ) \prod_{l=2}^{2|q|_{1}+|x|_{1}-1}
(l-\upalpha )\\
&\qquad \leq  (1-\upalpha ) \Upgamma (1-\upalpha  ) (2|q|_{1}+|x|_{1}-1)!
= \frac{(1-\upalpha ) \Upgamma (1-\upalpha  )}{2|q|_{1}+|x|_{1}}  (2|q|_{1}+|x|_{1})!\ .
\end{align*}
Inserting this bound in $ S_{2} (\upalpha, |x|_{1})$ we obtain
\begin{equation}\label{S2eq1}
 S_{2} (\upalpha, |x|_{1})\leq   (2d)^{\upalpha }(1-\upalpha ) \Upgamma (1-\upalpha  ) \sum_{|q|_{1}\geq 1}\frac{1}{2|q|_{1}+|x|_{1}}
\frac{(2|q|_{1}+|x|_{1})!}{\prod_{i=1}^{d} q_{i}!(q_{i}+|x_{i}|)!}\ \frac{1}{(2d)^{2|q|_{1}+|x|_{1}}}.
\end{equation}
In the case $d=1$ we have
\[
S_{2} (\upalpha, |x|_{1})\leq   (2d)^{\upalpha }(1-\upalpha ) \Upgamma (1-\upalpha  )  \sum_{q\geq 1}\frac{1}{2q+|x|}
\frac{1}{2^{2q+|x|}} \frac{(2q+|x|)!}{q!(q+|x|)!}.
\]
Note that  the binomial coefficient $\frac{n!}{n_{1}! n_{2}!}$ is maximal at  $n_{1}=n_{2},$ hence, 
 together with Stirling's formula, we get
\[
 \frac{n!}{n_{1}! n_{2}!}\leq \frac{n!}{\left(\lfloor{\frac{n}{2}\rfloor}\right)!^{2}}\leq C_{1}\ \frac{2^{n}}{n^{\frac{1}{2}}},
\]
for some constant $C_{1}>0.$ Inserting this bound above we obtain
\[
\sum_{q\geq 1}\frac{1}{2q+|x|}
\frac{1}{2^{2q+|x|}} \frac{(2q+|x|)!}{q!(q+|x|)!}\leq C_{1}  \sum_{q\geq 1}\frac{1}{2q+|x|}\frac{1}{2^{2q+|x|}} \frac{2^{2q+|x|}}{\sqrt{q}}
\leq  \frac{C_{1}}{2} \sum_{q\geq 1}
\frac{1}{q^{\frac{3}{2}}}=: C_{1}'<\infty,
\]
which proves \eqref{eq:claimS2} for $d=1.$\vspace{0,2cm}

In the case $d\geq 2$ we write
\begin{align*}
 S_{2} (\upalpha, |x|_{1})
&=   (2d)^{\upalpha }(1-\upalpha ) \Upgamma (1-\upalpha  )  \sum_{n\geq 1}\frac{1}{2n+|x|_{1}}
\frac{1}{(2d)^{2n+|x|_{1}}}  \sum_{|q|_{1}=n}\frac{(2n+|x|_{1})!}{\prod_{i=1}^{d} q_{i}!(q_{i}+|x_{i}|)!}
\end{align*}
We develop the binomial coefficient as follows
\begin{equation}\label{coeffbinomial}
\frac{(2n+|x|_{1})!}{\prod_{i=1}^{d} q_{i}!(q_{i}+|x_{i}|)!}= \frac{(2n+|x|_{1})!}{n! \, (n+|x|_{1})!} \ \frac{(n+|x|_{1})!}{\prod_{i=1}^{d} (q_{i}+|x_{i}|)!}\ \frac{n!}{\prod_{i=1}^{d} q_{i}!}.
\end{equation}
Using $\frac{n!}{q!(n-q)!}\leq 2^{n}$ for all $0\leq q\leq n,$ and the fact that the multinomial coefficient $\frac{n!}{\prod_{i=1}^{d}q_{j}!}$ is maximal when all $q_{j}$ are equal, together with Stirling's formula, we get
\[
 \frac{(2n+|x|_{1})!}{n! \, (n+|x|_{1})!} \leq 2^{2n+|x|_{1}}\quad \mbox{and}\quad 
 \frac{n!}{\prod_{i=1}^{d}q_{j}!}\leq \frac{n!}{\left(\lfloor{\frac{n}{d}\rfloor}\right)!^{d}}\leq C_{d} \frac{d^{n}}{n^{\frac{d-1}{2}}},
\]
for some constant $C_{d}>0.$  Inserting these bounds in \eqref{coeffbinomial} we obtain
\[
\frac{(2n+|x|_{1})!}{\prod_{i=1}^{d} q_{i}!(q_{i}+|x_{i}|)!}\leq
 2^{2n+|x|_{1}}  \frac{(n+|x|_{1})!}{\prod_{i=1}^{d} (q_{i}+|x_{i}|)!} C_{d} \frac{d^{n}}{n^{\frac{d-1}{2}}}.
\]
Using
\[
 \sum_{|q|_{1}=n} \frac{(n+|x|_{1})!}{\prod_{i=1}^{d} (q_{i}+|x_{i}|)!}\leq d^{n+|x|_{1}}
\]
and inserting all these bounds in $S_{2} (\upalpha, |x|_{1})$ we obtain 
\begin{align*}
 S_{2} (\upalpha, |x|_{1})&\leq  (2d)^{\upalpha } (1-\upalpha ) \Upgamma (1-\upalpha  )  C_{d} \sum_{n\geq 1}\frac{1}{2n+|x|_{1}}
 \frac{1}{(2d)^{2n+|x|_{1}}} 2^{2n+|x|_{1}}  d^{n+|x|_{1}} \frac{d^{n}}{n^{\frac{d-1}{2}}}\\
&\leq   (2d)^{\upalpha } (1-\upalpha ) \Upgamma (1-\upalpha  )  C_{d} \sum_{n\geq 1}\frac{1}{2 n^{1+\frac{d-1}{2}}}=
   (2d)^{\upalpha }(1-\upalpha ) \Upgamma (1-\upalpha  ) C'_{d}
\end{align*}
where $0<C'_{d}<\infty .$
This proves \eqref{eq:claimS2} for $d>1$.\vspace{0,2cm}

We study now the term $S_{1}(\upalpha, |x|_{1}).$
We distinguish three cases. 
\vspace{0,2cm}

\textit{Case 1: $|x|>1$}. In this case $|x|_{1}\geq 2$ and   therefore, using $0<\upalpha<1, $  
\[
\Upgamma (|x|_{1}-\upalpha  )\leq  (1-\upalpha ) \Upgamma (1-\upalpha  )  (|x|_{1}-1)!\ .
\]
It follows, using \eqref{defS1},
\[
\frac{S_{1} (\upalpha, |x|_{1})}{\abs{\Upgamma (-\upalpha)} }\leq
\frac{ (1-\upalpha )\Upgamma (1-\upalpha  )}{\abs{\Upgamma (-\upalpha)}  }
\frac{(|x|_{1}-1)! }{\prod_{i=1}^{d}|x_{i}|!}\ \frac{1}{(2d)^{|x|_{1}-\upalpha}}\leq
 \upalpha  (1-\upalpha  ) (2d)^{\upalpha }.
\]
Together with  \eqref{S2bound} this yields \eqref{bound1}.

\vspace{0,2cm}

\textit{Case 2: $|x|_{2}=1=|x|_{1}$}. In this case 
\[
 \frac{S_{1} (\upalpha ,1)}{\abs{\Upgamma (-\upalpha)}}=  \frac{1}{(2d)^{1-\upalpha }}  \frac{\Gamma (1-\upalpha)}{\abs{\Upgamma (-\upalpha)}}
= \frac{\upalpha }{(2d)^{1-\upalpha }}. 
\]
Together with  \eqref{S2bound} this yields \eqref{bound2}.
\vspace{0,2cm}

\textit{Case 3: $|x|_{2}=0=|x|_{1}$}. 
Inserting $ 1- e^{-2dt} =\int_{0}^{1} 2dt e^{-2dts} ds$ in \eqref{defS10} we obtain
\[
\frac{S_{1} (\upalpha,0 )}{\abs{\Upgamma (-\upalpha)}}= (2d)^{\upalpha }\frac{\Upgamma (1-\upalpha  )}{\abs{\Upgamma (-\upalpha)}  }\int_{0}^{1} \dd s\ s^{-1+\upalpha }=  (2d)^{\upalpha } \frac{\Upgamma (1-\upalpha  )}{\upalpha  \abs{\Upgamma (-\upalpha)}  }
=  (2d)^{\upalpha } .
\]
Hence $\lim_{\upalpha \to 1} \frac{S_{1} (\upalpha ,0)}{\abs{\Upgamma (-\upalpha)}}=2d$ and
$\lim_{\upalpha \to 1} \frac{S_{1} (\upalpha,0 )}{\abs{\Upgamma (-\upalpha)}}=1.$
Together with  \eqref{S2bound} this yields \eqref{bound3} 
and concludes the proof of the theorem.
\end{proof}

%%%%%%%%%%%%%%%%%%%%%%%%%%%%%%%%%%%%%%%%%%%%%%%%%%%%%%%%%%%%%%%%%%%%%

\subsection{Resolvent decay}

In this section we consider the operator  $[(-\Delta)^\upalpha+m^2]^{-1}$ with $m>0.$ This operator  is 
well defined and bounded since   $-m^2\notin\sigma((-\Delta)^{\upalpha})=[0,(4d)^\upalpha],$ for all $m>0.$

Recall that  (see \cite[Thm.~2.2]{gebert2020lipschitz} or \cite[Lemma~2.1]{Slade2018})
\begin{equation}\label{eq:calphadef}
c_{\upalpha,d}= \lim_{|x-y|\to \infty } |x-y|^{d+2\upalpha } ( - (-\Delta )^{\upalpha } (x,y))>0.
\end{equation}

\begin{theorem} \label{prop:decay-massive-resolv}
Set $m>0$ and $0<\upalpha <1.$ The matrix elements of the resolvent satisfy
\begin{equation}\label{eq:positivityG}
  \inf_{m>0}( (-\Delta )^{\upalpha }+m^{2})^{-1} (x,y)>0\qquad \forall x,y\in \Z^{d},
\end{equation}
and 
\begin{equation}\label{eq:limmpos}
\lim_{|x-y|\to \infty } |x-y|^{d+2\upalpha } ( (-\Delta )^{\upalpha }+m^{2})^{-1} (x,y)=\frac{c_{\upalpha,d}}{m^{4}}
\end{equation}
where $c_{\upalpha,d}$ is the constant introduced in \eqref{eq:calphadef}.
Moreover
there are  constants $C_{1}=C_{1} (m,\upalpha,d)>0$ and
$c_{1}=c_{1} (\upalpha,d)>0$ such that  
\begin{equation}\label{eq:massivedecay}
 \frac{c_{1}}{|x-y|^{d+2\upalpha }} \leq  ((-\Delta)^{\upalpha }+m^{2})^{-1} (x,y)\leq \frac{C_{1}}{|x-y|^{d+2\upalpha }}\qquad
\forall x\neq y.
\end{equation}
where the constant $c_{1}$ is independent of the mass $m.$
 \end{theorem}
%%%%%%%%%%%%%%%%%%%%%%%%%%%%%%%
Note that the asymptotic behavior \eqref{eq:limmpos} is compatible with the upper bound obtained in \cite[Lemma~3.2]{Slade2018})
with other techniques.

\begin{proof}
To prove the lower bound note that 
\[
(-\Delta )^{\upalpha }+m^{2}= m_{\upalpha }^{2}\mathrm{Id} -P,
\]
where $P (x,y):=-(-\Delta )^{\upalpha }(x,y)=|(-\Delta )^{\upalpha }(x,y)|>0$ for $x\neq y,$ $P (x,x):=0$ and
$ m_{\upalpha }^{2}=m^{2}+ (-\Delta)^{\upalpha } (0,0).$
For  $m>0$ the Neumann series
\[
( (-\Delta )^{\upalpha }+m^{2})^{-1} (x,y)  = \frac{1}{ m_{\upalpha }^{2}}\mathrm{Id}+ \sum_{n\geq 1} \frac{1}{ m_{\upalpha }^{2}} \left(  P\frac{1}{ m_{\upalpha }^{2}}\right)^{n} (x,y)
\]
is a sum of positive terms and converges for all $x,y\in \Z^{d}$. Bounding the sum below by the first non-zero term
we obtain for $x=y$
\begin{equation}\label{eq:boundGm00}
( (-\Delta )^{\upalpha }+m^{2})^{-1} (x,x)\geq  \frac{1}{ m_{\upalpha }^{2}}\geq  \frac{1}{(-\Delta)^{\upalpha } (0,0)}>0
\end{equation}
uniformly in $m>0.$
In the case   $x\neq y,$   using also \eqref{Eq.decay}, we obtain
\begin{align}
( (-\Delta )^{\upalpha }+m^{2})^{-1} (x,y)&\geq  \frac{1}{ m_{\upalpha }^{4}}P (x,y)=
 \frac{1}{ m_{\upalpha }^{4}}| (-\Delta )^{\upalpha }(x,y)|\geq
 \frac{c_{\upalpha,d}}{ m_{\upalpha }^{4}} \frac{1}{\abs{x-y}^{d+2\upalpha}}\nonumber\\
&\geq
 \frac{c_{\upalpha,d}}{  (-\Delta)^{\upalpha } (0,0)^{2}} \frac{1}{\abs{x-y}^{d+2\upalpha}}.
\label{eq:lowerbound}\end{align}
This concludes the proof of \eqref{eq:positivityG} and the lower bound in \eqref{eq:massivedecay}.\vspace{0,2cm}

To prove \eqref{eq:limmpos} and the  upper bound in \eqref{eq:massivedecay},  note that  $\qty[(-\Delta)^\upalpha+m^2]^{-1}$ is defined via discrete Fourier transform as follows:
\begin{align}\label{FT.Inv.FL}
    \qty[(-\Delta)^\upalpha+m^2]^{-1}(x,y)&=\int_{[-\uppi,\uppi]^{d}}\frac{\dd k}{(2\uppi)^{d}} \frac{e^{i(x-y)\cdot k}}{\mathrm{f} (k)^\upalpha+m^2},
\end{align}
where 
\begin{equation}\label{eq:fdef}
\mathrm{f} (k):=\sum_{j=1}^{d}2(1-\cos k_j).
\end{equation}
This operator is invariant under translations, hence it suffices to consider the case $y=0.$ Applying $N\geq 2$ times  the
identity 
\[
\frac{1}{\mathrm{f} (k)^\upalpha+m^2}=\frac{1}{m^{2}}-
\frac{\mathrm{f} (k)^\upalpha}{m^{2} ( \mathrm{f} (k)^\upalpha+m^2 )}
\]
we obtain
\[
\frac{1}{\mathrm{f} (k)^\upalpha+m^2}=\sum_{j=0}^{N-1} \frac{(-1)^{j}}{m^{2 (j+1)}}\mathrm{f} (k)^{\upalpha j}+
\frac{(-1)^{j}}{m^{2N}} \frac{\mathrm{f} (k)^{N\upalpha}}{\mathrm{f} (k)^\upalpha+m^2}.
\]
Inserting this decomposition in the integral and using  $\int_{[-\uppi,\uppi]^{d}} e^{ix\cdot k}\dd k=0$
for all $x\neq 0,$ we obtain
\begin{align*}
\qty[(-\Delta)^\upalpha+m^2]^{-1}(x,0)&=\frac{1}{m^{4}} (-(-\Delta)^{\upalpha} (x,y))+\sum_{j=2}^{N-1}\tfrac{(-1)^{j}}{m^{2 (j+1)}}(-\Delta)^{j\upalpha} (x,0)\\
&+\tfrac{ (-1)^{N}}{m^{2N}(2\uppi)^{d}} \int_{[-\uppi,\uppi]^{d}}
\dd k\ F (\mathrm{f} (k)^\upalpha)\ e^{ix\cdot k},
\end{align*}
where the function $F\colon [0,\infty )\to  [0,\infty )$ is  defined by
\[
F(x):=  \frac{x^{N}}{x+m^2}.
\]
Since $0<\upalpha <1$ there is a
$N_{\upalpha}\geq 2$ such that $\upalpha( N_{\upalpha}-1) \leq 1$ and $\upalpha N_{\upalpha} >1.$ Setting $N=N_{\upalpha}$  and using \eqref{Eq.decay}
(for $\upalpha(N_{\upalpha}-1)=1$ we obtain $(-\Delta) (x,y)$ which is a finite range kernel)
\[
\left | \sum_{j=2}^{N_{\upalpha}-1}\frac{(-1)^{j}}{m^{2 (j+1)}}[(-\Delta)^{j\upalpha} (x,y)]\right |\leq
  \sum_{j=2}^{N_{\upalpha}-1}\frac{C_{j\upalpha,d}}{m^{2 (j+1)}\abs{x-y}^{d+2j\upalpha}}<  \frac{C_{1}^{(1)}}{\abs{x-y}^{d+4\upalpha}},
\]
where
\[
C_{1}^{(1)}:=N_{\upalpha}\max_{j=2\dotsc N_{\upalpha} } \frac{C_{j\upalpha,d}}{m^{2 (j+1)}}.
\]
The limit \eqref{eq:limmpos} and the  upper bound in \eqref{eq:massivedecay} now follow from the following estimate
\begin{equation}\label{eq:boundF}
\left|\int_{[-\uppi,\uppi]^{d}}
\dd k\ F (\mathrm{f} (k)^\upalpha)\ e^{ix\cdot k}\right |\leq \frac{C_{1}^{(2)}}{|x|^{d+2}}\quad
\forall x\neq 0,
\end{equation}
for some constant $C_{1}^{(2)}>0.$ To prove it, note that $F\in C^{\infty } ([0,\infty ))$ with $F (x)=O (x^{N})$ as $x\to 0.$
On the contrary, the function $k\mapsto \mathrm{f} (k)^\upalpha$ is in  $ C^{\infty } ([-\pi,\pi  ]^{d}\setminus \{0 \}).$
The first derivative equals
\[
\partial_{k_{j}}\mathrm{f} (k)^\upalpha=\upalpha \mathrm{f} (k)^\upalpha \ \frac{2\sin k_{j}}{\mathrm{f} (k)},
\]
and hence $|\partial_{k_{j}}\mathrm{f} (k)^\upalpha|\leq O (|k|^{2\upalpha-1 }).$ Any additional derivative brings 
an additional $|k|^{-1}$ divergence factor. Therefore near $k=0$ we have
\begin{equation}\label{eq:asymptF}
|\partial^{\beta }_{k}F (\mathrm{f} (k)^\upalpha)|\leq C_{|\beta |,\upalpha,N} O (|k|^{2\upalpha N_{\upalpha } -|\beta |}). 
\end{equation}
This implies that $\partial^{\beta }_{k}F (\mathrm{f} (k)^\upalpha)\in L^{1} ([-\pi,\pi  ]^{d})$ for all $|\beta |\leq d+2.$
In addition $\mathrm{f} (k) $ is periodic  with period $2\pi $  in all variables.
Since $\mathrm{f} (k) $ is even, we can assume without loss of generality $x_{j}\geq 0$
$\forall j=1,\dotsc, d,$ so that $|x|_{1}=\sum_{j}x_{j}.$
We argue
\begin{align*}
-i x_{j}\int_{[-\uppi,\uppi]^{d}} \dd k\ \partial_{k}^{\beta }F (\mathrm{f} (k)^\upalpha))\ e^{ix\cdot k}&=
-\int_{[-\uppi,\uppi]^{d}} \dd k\ \partial_{k}^{\beta }F (\mathrm{f} (k)^\upalpha))\ \partial_{k_{j}} e^{ix\cdot k}\\
&
= -\lim_{\varepsilon \to 0}\int_{[-\uppi,\uppi]^{d}\setminus B_{\varepsilon } (0)} \dd k\ \partial_{k}^{\beta }F (\mathrm{f} (k)^\upalpha))\ \partial_{k_{j}} e^{ix\cdot k}
\end{align*}
Performing partial integration we obtain
\begin{align*}
\int_{[-\uppi,\uppi]^{d}\setminus B_{\varepsilon } (0)} \dd k\ \partial_{k}^{\beta }F (\mathrm{f} (k)^\upalpha))\ \partial_{k_{j}} e^{ix\cdot k}&=
\int_{\partial B_{\varepsilon } (0)}
\dd \mathcal{H}^{d-1} \partial_{k}^{\beta } F (\mathrm{f} (k)^\upalpha))\ 
\nu_{j} (k) e^{ix\cdot k}\\
&- \int_{[-\uppi,\uppi]^{d}\setminus B_{\varepsilon } (0) } \dd k\
\partial_{k_{j}}\partial_{k}^{\beta }F (\mathrm{f} (k)^\upalpha))\ e^{ix\cdot k}  \
\end{align*}
where $\nu (k):=\frac{1}{|k|}k$ and the periodicity of $\mathrm{f} (k)$ garantees there is no contribution from the boundary of $[-\pi,\pi ]^{d}.$
For $|\beta |\leq d+1$ and $\upalpha N_{\upalpha } >1,$ using  \eqref{eq:asymptF}
\[
 \limsup_{\varepsilon \to 0}|\int_{\partial B_{\varepsilon } (0)}
\dd \mathcal{H}^{d-1}  \partial_{k}^{\beta } F (\mathrm{f} (k)^\upalpha))\ \nu_{j} (k) e^{ix\cdot k} |\leq
C  \limsup_{\varepsilon \to 0} \varepsilon^{2\upalpha N_{\upalpha }-|\beta |} \varepsilon^{d-1}=0,
\]
and hence
\[
-i x_{j}\int_{[-\uppi,\uppi]^{d}} \dd k\ \partial_{k}^{\beta }F (\mathrm{f} (k)^\upalpha))\ e^{ix\cdot k}=
 \int_{[-\uppi,\uppi]^{d} } \dd k\
\partial_{k_{j}}\partial_{k}^{\beta }F (\mathrm{f} (k)^\upalpha))\ e^{ix\cdot k} ,
\]
where the last integral is well defined since  $|\beta |+1\leq d+2.$
The integrability of the derivative ensures we can
 repeat the procedure above inductively until $|\beta |+1=d+2.$ This concludes the proof of \eqref{eq:boundF} and of the theorem.
\end{proof}
%%%%%%%%%%%%%%%%%%%%%%%%%%%%%%%%%%%%%

\subsection{Inverse}
 
    \begin{theorem} \label{decayGreenszero}
        Let $0<\upalpha < \frac{d}{2}.$ The inverse $(-\Delta)^{-\upalpha}(x,y):=\lim_{m\downarrow 0}\qty[(-\Delta)^\upalpha+m^2]^{-1}(x,y)$  is well-defined and admits the represention via discrete Fourier transform
\begin{equation}\label{def.invFL}
(-\Delta)^{-\upalpha}(x,y) =\int_{[-\uppi,\uppi]^{d}}\frac{\dd k}{(2\uppi)^{d}} \frac{e^{i(x-y)\cdot k}}{\mathrm{f} (k)^\upalpha},
\end{equation}
This operator is invariant under translations, its  matrix elements satisfy
$ (-\Delta)^{-\upalpha }(x,y)>0$ 
$\forall x, y\in \mathbb{Z}^{d},$  and 
\begin{equation}\label{eq:limmposm=0}
\lim_{|x-y|\to \infty } |x-y|^{d-2\upalpha } (-\Delta)^{-\upalpha }(x,y)=\mathfrak{c}_\upalpha
\end{equation}
where $\mathfrak{c}_\upalpha$ is the constant introduced in \eqref{calphadef}. Moreover there are  constants $C_{2}=C_{2} (\upalpha,d)>0$ and
$c_{2}=c_{2} (\upalpha,d)>0$ such that  
 \begin{align}\label{Ineq.Inv.FL}
c_{2}\, \frac{1}{|x-y|^{d-2\upalpha }}\leq (-\Delta)^{-\upalpha} (x,y)\leq
C_{2}\,  \frac{1}{|x-y|^{d-2\upalpha }}\qquad \forall x\neq y.
\end{align}

\end{theorem}
%%%%%%%%%%%%%%%%%%%%%%%%%%%%%%%
 The fact that $(-\Delta)^{-\upalpha}$ is well-defined and Ineq. \ref{Ineq.Inv.FL} holds are known (see e.g. 
  \cite[Sect.~2]{Slade2018} and references therein). Here, we provide an alternative,
 more analytical proof, which we believe is new in this context.
 It uses  the discrete Fourier transform and is based on arguments in \cite[Lemma A.1]{gebert2020lipschitz}.

%%%%%%%%%%%%%%%%%%%%%%%%%%%%%%%%%%%%%%%%%%%%%%%%%
\begin{proof}\hspace{2cm}

By  \eqref{eq:positivityG} we have $(-\Delta)^{-\upalpha}(x,y):=\lim_{m\downarrow 0}\qty[(-\Delta)^\upalpha+m^2]^{-1}(x,y)>0$ $\forall x,y\in \mathbb{Z}^{d}.$ 
Moreover, remember that, for all $m>0,$  (cf. \eqref{FT.Inv.FL})
\[
 \qty[(-\Delta)^\upalpha+m^2]^{-1}(x,y)=\int_{[-\uppi,\uppi]^{d}}\frac{\dd k}{(2\uppi)^{d}} \frac{e^{i(x-y)\cdot k}}{\mathrm{f} (k)^\upalpha+m^2},
\]
where $\mathrm{f} (k)$ is defined in \eqref{eq:fdef}.
Note that $[\mathrm{f} (k)^\upalpha+m^2]^{-1}\leq \mathrm{f} (k)^{-\upalpha}$ which is an unbounded integrable function.
Indeed this function behaves near  $k=0$ as $\frac{1}{|k|^{2\upalpha }}$ which  integrably divergent
as long as $\upalpha<\frac{d}{2}.$ Therefore, by dominated convergence, the limit $\varepsilon \to 0$ is well defined
and formula \eqref{def.invFL} holds.\vspace{0,2cm}

To prove  \eqref{Ineq.Inv.FL} we approximate the discrete  Laplacian  $-\Delta $ on $\Z^{d},$ with eigenvalues
$\mathrm{f} (k),$ by the continuous 
Laplacian $-\Delta_{c}$  on $\R^{d},$ with eigenvalues $|k|^{2},$ and use the known decay $|(-\Delta_{c})^{-1} (x,y)|\leq \frac{C}{|x-y|^{d-2\upalpha }},$
which holds in distributional sense for some $C>0$ (cf. Proposition \ref{Prop.2.Theo1.6} below for a
precise statement). By translation invariance
it suffices to consider the case $y=0,$ $x\neq 0. $

The two functions $\mathrm{f} (k)$ and $|k|^{2}$ coincide only near $k=0,$ therefore we introduce a smooth cut-off function
$\uppsi\in C^\infty_{c}(\R^{d}, [0,1])$ such that $\mathrm{supp}\uppsi\subset  B_{1}(0)$ and $ \uppsi(k)=1$
$\forall k\in B_{\frac{1}{2}}\qty(0).$ Hence $ (-\Delta)^{-\upalpha}(0,x)= I_{1} (x)+ I_{2} (x)$ where 
\begin{equation}
I_{1} (x):= \int_{[-\uppi,\uppi]^{d}}\frac{\dd k}{(2\uppi)^{d}} \frac{\uppsi (k)}{\mathrm{f} (k)^\upalpha}e^{ix\cdot k},
\qquad 
I_{2} (x):= \int_{[-\uppi,\uppi]^{d}}\frac{\dd k}{(2\uppi)^{d}}  \frac{1-\uppsi (k)}{\mathrm{f} (k)^\upalpha}e^{ix\cdot k}.
\end{equation}
Note that $k\mapsto  \frac{1-\uppsi (k)}{\mathrm{f} (k)^\upalpha}\in C^{\infty} ([-\pi,\pi ]^{d}).$  Moreover,
along the boundary of $[-\uppi,\uppi]^{d}$  we have
$ \frac{1-\uppsi (k)}{\mathrm{f} (k)^\upalpha}= \frac{1}{\mathrm{f} (k)^\upalpha},$ which is a periodic function with period $2\pi $
in all variables. Therefore, by partial integration, we obtain
\[
|x_{j}^{N}I_{2} (x)|\leq  \int_{[-\uppi,\uppi]^{d}}\frac{\dd k}{(2\uppi)^{d}}\left | \partial_{k_{j}}^{N} \frac{1-\uppsi (k)}{\mathrm{f} (k)^\upalpha}\right|\leq C_{N},
\]
where  no boundary contribution appears by periodicity. It follows, for all $N\geq 1$
\[
|I_{2} (x) |\leq \frac{C_{N}}{|x|^{N}}\qquad \forall x\neq 0,
\]
for some constant $C_{N}>0.$ Therefore we only need to study the first integral $I_{1} (x).$
We write
\[
I_{1} (x)= \int_{[-\uppi,\uppi]^{d}}\frac{\dd k}{(2\uppi)^{d}}
\frac{1}{\abs{k}^{2\upalpha}}\Phi(k)\ e^{ix\cdot k}=
 \int_{\R^{d}}\frac{\dd k}{(2\uppi)^{d}}
\frac{1}{\abs{k}^{2\upalpha}}\Phi(k)\ e^{ix\cdot k}
\]
where we defined
\begin{align}\label{def.Phi}
    \Phi(k):=\begin{cases}\qty(\frac{\abs{k}^{2}}{\mathbf{\mathrm{f}}(k)})^{\upalpha}\uppsi(k),\qquad& k\neq 0,\\
    1,\qquad& k= 0.
    \end{cases}
\end{align}
and in the last step we used the fact the $\uppsi$ has support inside $B_{1} (0)$ to extend the integral from $[-\uppi,\uppi]^{d}$ to $\R^{d}.$ The function $\Phi $ is smooth $\Phi(k)\in C_c^{\infty}(\R^{d})\subset\mathcal{S}(\R^{d}),$ and hence  it is the continuous Fourier transform of a function $\upvarphi\in \mathcal{S}(\R^{d})$
(see \cite[Corollary 2.2.15]{loukas2014classical}) 
\[
\Phi (k) =\hat{\upvarphi} (k):=
\frac{1}{(2\uppi)^{\frac{d}{2}}}\
\int_{\R^{d}}\dd y\:e^{-iy\cdot k}\upvarphi(y).
\]
It follows, by Proposition  \ref{Prop.2.Theo1.6} below,
\[
I_{1} (x)=\int_{\R^{d}}\frac{\dd k}{(2\uppi)^{d}}
\frac{1}{\abs{k}^{2\upalpha}}\Phi(k)\ e^{ix\cdot k}=\mathfrak{c}_{\upalpha } \int_{\R^{d}}\dd y \ \frac{1}{|x-y|^{d-2\upalpha }}
\upvarphi (y)
\]
where the constant $\mathfrak{c}_\upalpha$ is given in \eqref{calphadef}.
The integral above is well-defined since
$\upvarphi\in\mathcal{S}(\R^{d})$\footnote{
The function $\mathrm{I}_\upalpha [\upvarphi](y):= \mathfrak{c}_\upalpha\int_{\R^{d}}\dd y \ \frac{1}{|x-y|^{d-2\upalpha }} \upvarphi (y)$ is called Riesz potential. }.
Using this result and Proposition \ref{Prop.3.Theo1.6} below,  we argue
\begin{align*}
 \lim_{\abs{x}\rightarrow\infty}\abs{x}^{d-2\upalpha}(-\Delta)^{-\upalpha}(0,x)&= \mathfrak{c}_\upalpha
 \lim_{\abs{x}\rightarrow\infty} \int_{\R^{d}}\dd y \ \frac{\abs{x}^{d-2\upalpha}}{|x-y|^{d-2\upalpha }} \upvarphi (y)\\
&= \mathfrak{c}_\upalpha \int_{\R^{d}}\dd y\  \upvarphi (y)= \mathfrak{c}_\upalpha\hat\upvarphi (0)=  \mathfrak{c}_\upalpha\Phi(0)= \mathfrak{c}_\upalpha>0,
\end{align*}
where in the last step we used $\Phi(0)=1.$ 
The  limit \eqref{eq:limmposm=0}, as well as the upper and lower bounds in \eqref{Ineq.Inv.FL} now follow.
\end{proof}
%%%%%%%%%%%%%%%%%%%%%%%%%%%%%%%%%%%%%%%%%%%%%%%%%%%5

We collect finally two techical results that are necessary for the proof above.
The first  proposition can be found 
in \cite[Section~5.9]{lieb2001analysis} (the constants are slightly different because of our choice of definition of Fourier transform), see also \cite[Chapter~5]{stein-book}.  For completeness we give here a sketch of the proof.
The second proposition is based on the same arguments as the proof of \cite[Lemma~A.1]{gebert2020lipschitz}.
%%%%%%%%%%%%%%%%%%%%%%%%%%%%%%%%%%%%%%%%%
\begin{proposition}\textrm{\cite[Thm.~5.9]{lieb2001analysis}}\label{Prop.2.Theo1.6}
    Let $\upalpha\in (0,\frac{d}{2})$ and  
let  $\upvarphi\in \mathcal{S}(\R^{d})$, then 
    \begin{align}
     \int_{\R^{d}}  \frac{\dd k}{(2\uppi)^{\frac{d}{2}}}\frac{1}{\abs{k}^{2\upalpha}}\hat{\upvarphi}(k) e^{+ik\cdot x}
=  \mathfrak{c}_\upalpha\int_{\R^{d}}\dd y \ \frac{1}{|x-y|^{d-2\upalpha }} \upvarphi (y)
\label{eq.Lieb}
    \end{align}
    where
    \begin{equation}\label{calphadef}
\mathfrak{c}_{\upalpha }:=\frac{\Upgamma(\frac{d}{2}-\upalpha)}{\Upgamma(\upalpha)}2^{\frac{d}{2}-2\upalpha}.
\end{equation}
\end{proposition}
%%%%%%%%%%%%%%%%%%%%%%%%%%%%%%%%%%%%%%%%%%%%%%%
\begin{proof} Using the identity, which holds for all $\rho >0$ and $0<\upalpha <1,$
\[
    \frac{1}{\uprho^\upalpha}=\frac{1}{2^\upalpha\Upgamma(\upalpha)}\int_0^\infty \dd t \:t^{\upalpha-1}e^{-\frac{t}{2}\uprho },
\]
 we can write
\begin{align}\label{IntRepPower}
    \frac{1}{\abs{k}^{2\upalpha}}=\frac{1}{2^\upalpha\Upgamma(\upalpha)}\int_0^\infty\dd t \: t^{\upalpha-1}e^{-\frac{t}{2}\abs{k}^2 },\qquad\forall k\in\R^{d}\setminus\qty{0}.
\end{align}
It follows, using Fubini and the Fourier transform of a product,
\begin{align*}
&   \int_{\R^{d}}  \frac{\dd k}{(2\uppi)^{\frac{d}{2}}}\frac{1}{\abs{k}^{2\upalpha}}\hat{\upvarphi}(k) e^{+ik\cdot x}=
\frac{1}{2^\upalpha\Upgamma(\upalpha)}
\int_0^\infty\dd t \: t^{\upalpha-1}\int_{\R^{d}}  \frac{\dd k}{(2\uppi)^{\frac{d}{2}}}e^{-\frac{t}{2}\abs{k}^2 }\hat{\upvarphi}(k) e^{+ik\cdot x}\\
&\qquad =\frac{1}{2^\upalpha\Upgamma(\upalpha)}
\int_0^\infty\dd t \: t^{\upalpha-1}\frac{1}{t^{\frac{d}{2}}}
\int_{\R^{d}} \dd y e^{-\frac{1}{2t}\abs{x-y}^2 }\upvarphi(y) \\
&\qquad =
\frac{1}{2^\upalpha\Upgamma(\upalpha)}
\int_{\R^{d}} \dd y \upvarphi(y)
\int_0^\infty\dd t \: t^{\upalpha-1}\frac{1}{t^{\frac{d}{2}}} e^{-\frac{1}{2t}\abs{x-y}^2 }
=\mathfrak{c}_{\upalpha }\int_{\R^{d}}\dd y \ \frac{1}{|x-y|^{d-2\upalpha }} \upvarphi (y).
\end{align*}
This concludes the proof.
\end{proof}
%%%%%%%%%%%%%%%%%%%%%%%%%%%%%%%%%%%%%%

 %%%%%%%%%%%%%%%%%%%%%%%%%%%%%%%%%%%%%%%%%%%%%%%%%%%%
\begin{proposition}\label{Prop.3.Theo1.6} Let $\upalpha\in (0,\frac{d}{2})$ and  
let  $\upvarphi\in \mathcal{S}(\R^{d})$, then
\begin{align}
    \lim_{\abs{x}\rightarrow \infty}\int_{\R^{d}}\dd y \ \frac{|x|^{d-2\upalpha}}{|x-y|^{d-2\upalpha }} \upvarphi (y)= \int_{\R^{d}}\dd y \ \upvarphi (y). \label{eq.LiebRiesz}
\end{align}
\end{proposition}
%%%%%%%%%%%%%%%%%%%%%%%%%%%%%%%%%%%%%%%%%%%%%%%%%%%%%%%%%%
\begin{proof}

We decompose the integral as follows:
\[
\int_{\R^{d}}\dd y \ \frac{|x|^{d-2\upalpha}}{|x-y|^{d-2\upalpha }} \upvarphi (y)=
\int_{B_{\frac{|x|}{2}} (x)}\dd y \ \frac{|x|^{d-2\upalpha}}{|x-y|^{d-2\upalpha }} \upvarphi (y)+
\int_{B_{\frac{|x|}{2}}^{c} (x)}\dd y \ \frac{|x|^{d-2\upalpha}}{|x-y|^{d-2\upalpha }} \upvarphi (y).
\]
The first integral can be reorganized as
\[
\int_{B_{\frac{|x|}{2}} (x)}\dd y \ \frac{|x|^{d-2\upalpha}}{|x-y|^{d-2\upalpha }} \upvarphi (y)=
\int_{B_{\frac{|x|}{2}} (0)}\dd y \ \frac{|x|^{d-2\upalpha}}{|y|^{d-2\upalpha }} \upvarphi (x+y).
\]
Since 
 $\upvarphi\in \mathcal{S}(\R^{d})$ we have
 $|\upvarphi (y)|\leq \frac{C_{N}}{|y|^{N}}$ for $y\neq 0$ for all $N\geq 1.$
Therefore, since $|x+y|\geq \frac{|x|}{2}$ inside the ball $B_{\frac{|x|}{2}} (0)$ we have
\[
|\upvarphi (x+y)|\leq \frac{C_{N}}{|x+y|^{N}}\leq \frac{2^{N}C_{N}}{|x|^{N}}.
\]
Inserting this bound in the integral and fixing $N>d$ we obtain
\begin{align*}
\limsup_{|x|\to \infty } \int_{B_{\frac{|x|}{2}} (0)}\dd y \ \frac{|x|^{d-2\upalpha}}{|y|^{d-2\upalpha }} |\upvarphi (y+x)|&\leq \limsup_{|x|\to \infty } 2^{N}C_{N}\frac{|x|^{d-2\upalpha}}{|x|^{N}} \int_{B_{\frac{|x|}{2}} (0)}\dd y \ \frac{1}{|y|^{d-2\upalpha }}\\
&=\frac{2^{N}C_{N}|\mathcal{S}^{d-1}|}{2\upalpha\, 2^{2\upalpha }}\limsup_{|x|\to \infty } \frac{|x|^{d}}{|x|^{N}} =0,
\end{align*}
where $|\mathcal{S}^{d-1}|$ is the surface volume of the unit sphere in $\mathbb{R}^{d}$. Hence 
\[
\lim_{|x|\to \infty }\int_{B_{\frac{|x|}{2}} (x)}\dd y \ \frac{|x|^{d-2\upalpha}}{|x-y|^{d-2\upalpha }} \upvarphi (y)=0.
\]
We consider now the first integral. Note that, since, the center of the ball $B_{\frac{|x|}{2}} (x)$ escapes
at infinity as $|x|\to \infty $ it holds
\[
\lim_{|x|\to \infty}  \frac{|x|^{d-2\upalpha}}{|x-y|^{d-2\upalpha }}\mathbf{1}_{B_{\frac{|x|}{2}}^{c} (x)} (y)
=1\qquad \forall y\in\Z^{d}.
\]
Since  in addition
\[
 \frac{|x|^{d-2\upalpha}}{|x-y|^{d-2\upalpha }}\mathbf{1}_{B_{\frac{|x|}{2}}^{c} (x)} (y) |\upvarphi (y)|
 \leq 2^{d-2\upalpha }|\varphi (y)|\in L^{1} (\R^{d})
\]
holds, we obtain, by dominated convergence,
\[
\lim_{|x|\to \infty }\int_{B_{\frac{|x|}{2}}^{c} (x)}\dd y \ \frac{|x|^{d-2\upalpha}}{|x-y|^{d-2\upalpha }} \upvarphi (y)=\int_{\R^{d}}\dd y\ \upvarphi (y),
\]
which concludes the proof of the proposition.
\end{proof}\vspace{0,2cm}

%\end{appendices}

\paragraph{Acknowledgements} We would like to thank the anonymous
referee for carefully reading our manuscript and for suggestions that led to an improvement of the article.
We are grateful to Deutsche Forschungsgemeinschaft (\textit{GZ: RO 5965/1-1, AOBJ: 661454} \& \textit{Project-ID} 211504053 - \textit{SFB} 1060) and ANR RAW \textit{ANR-20-CE40-0012-01} for financial support.
R.M.E. was partially supported by the Global Math Network (Bonn) and acknowledges the hospitality of
Laboratoire AGM \& CY Cergy Paris Universit\'e where part of this work has been carried out.
C.R.-M.  acknowledges the hospitality of  Pontifical Catholic University of Santiago, Chile,
where part of this work was done. The authors
would like to thank M.~Gebert and M.~Heydenreich for references
(\cite{lieb2001analysis} and \cite{chen2015critical}), and R.~Matos for motivating the discussion around
\cite{schenker2015large}.

\paragraph{Data availability} Data sharing is not applicable to this article as no datasets were generated or analysed during the current study.

\section*{Declarations}

\paragraph{Conflict of interest} The authors have no competing interests to declare that are relevant to the content of this article.

\bibliographystyle{alpha}

%\bibliography{./fractional-laplacian}

\newcommand{\etalchar}[1]{$^{#1}$}

\end{document}